\documentclass[hidelinks,onefignum,onetabnum]{siamart251216}

\usepackage{amssymb,amsfonts}
\usepackage{mathtools}
\usepackage{booktabs}
\usepackage{graphicx}

\newsiamremark{assumption}{Assumption}
\newsiamremark{remark}{Remark}

\DeclareMathOperator{\Tr}{Tr}

\DeclareMathOperator*{\argmin}{arg\,min}
\DeclareMathOperator*{\argmax}{arg\,max}

\newcommand{\C}{\mathbb C}
\newcommand{\R}{\mathbb R}

\newcommand{\Hcal}{\mathcal H}
\newcommand{\Lcal}{\mathcal L}

\newcommand{\ip}[2]{\left\langle #1,#2\right\rangle}
\newcommand{\norm}[1]{\left\lVert #1\right\rVert}
\newcommand{\abs}[1]{\left\lvert #1\right\rvert}

\title{Task-Directed Spectral Design for Local Quantum Tomography%
\thanks{Submitted to the editors in August 2026.}}

\author{Guy Hay\thanks{School of Mathematical Sciences, Tel Aviv University,
Tel Aviv, Israel.}
\and Nir Sharon\footnotemark[2]}

\headers{Task-Directed Quantum Tomography}{G. Hay and N. Sharon}

\ifpdf
\hypersetup{
  pdftitle={Task-Directed Spectral Design for Local Quantum Tomography},
  pdfauthor={Guy Hay and Nir Sharon}
}
\fi

\begin{document}

\maketitle

\begin{abstract}
Many tomography tasks require only a low-dimensional function of the state. We study local pure-state tomography under projective-fidelity access when a parametric model, a credible center, and a differentiable downstream task are known before acquisition.  The pullback Fubini--Study metric and the task metric define coordinate-invariant generalized eigendirections, which we realize as a fixed library of geodesic fidelity probes.  We prove two optimality results: the probe aligned with a task coordinate is its unique variance-minimizing location up to orientation, and, when prior and measurement information are diagonal in this basis, a diminishing-return greedy rule exactly minimizes the integer-budget local task-variance surrogate over the spectral library. Under common per-shot information and the stated large-budget condition, the absolute and relative gaps to arbitrary probe directions are respectively \(O(B^{-3})\) and \(O(B^{-2})\). A closed-form local estimator is followed by a task-metric localization projection that cannot increase quadratic task loss when localization holds. In six-qubit numerical experiments, the Schmidt-spectrum design activates only one or two settings and has lower mean task error than the post hoc best adaptive results across all tested dimensions and copy budgets.  For full-state tomography, it likewise has lower mean infidelity throughout the tested local regime. The fixed batch uses no feedback loops and far fewer settings; under the illustrative latency model, its acquisition cost is roughly three to four orders of magnitude lower than that of the adaptive methods.
\end{abstract}




\section{Introduction}

Quantum tomography is usually posed as uniform reconstruction of an unknown
state.  That objective can be wasteful when the requested output is lower
dimensional, such as an entanglement spectrum or another differentiable
function of the state.  A fixed-batch protocol can further reduce acquisition
cost by using few proposal states and eliminating measurement--classical
feedback.  We consider the local setting in which a calibrated pure-state
model, a credible centre, and a downstream task are known before acquisition.
The design question is then: which state directions should be measured for
that task, and how should an integer copy budget be divided among them?

A normalized model \(\theta\mapsto\psi(\theta)\) induces the pullback
Fubini--Study metric \(G\), which measures physical state displacement.  The
differential of the requested output induces a task metric \(Q\).  The
generalized eigenproblem
\[
Qv_j=\lambda_jGv_j
\]
orders tangent directions by task sensitivity per unit state displacement.
We lift these directions to projective Hilbert space and realize them as
geodesic fidelity probes.  Their finite-shot Fisher information is rank one;
for direct projective readout its nonzero value is four per shot, independent
of the probe radius.  Two optimality theorems drive the design.  With prior
precision diagonal in the task basis, the probe located along \(\pm v_j\) is
the unique variance-minimizing location for task coordinate \(b_j\).  Over
these spectral probes, the task-variance surrogate separates into decreasing
marginal gains, and selecting the largest remaining gain gives its exact global
integer optimum for every budget.  Mixed probe directions can improve the
finite-budget joint design, but we prove that their advantage is only
\(O(B^{-3})\) in absolute risk and \(O(B^{-2})\) relatively under the stated
large-budget condition.

All settings and shot counts are fixed before outcomes are observed.  Scalar
geodesic inversions and prior shrinkage give a local estimate, and a task-metric
projection cannot increase its quadratic loss when the truth satisfies the
localization assumption.  Under the common fidelity-to-proposal access model,
our fixed design has acquisition cost \(Bs+K_+a\).  Our adaptive comparators are
self-guided quantum tomography (SGQT) \cite{ferrie_sgqt_2014}, orthogonalised
self-guided quantum tomography (OSGQT) \cite{dekkers_osgqt_2026}, and practical
adaptive quantum tomography (PAQT) \cite{granade_paqt_2016}.  Each uses two
settings and one feedback round per iteration, giving acquisition cost
\(Bs+2Ta+Tr\).
In the reported comparisons, our design uses one or two task settings, or 6
and 15 full-state settings, and therefore costs less whenever setting or
feedback has nonzero cost.  Under the illustrative
\(s:a:r=1:10000:400000\) scenario, its acquisition cost is lower by factors of
\(687\)--\(7434\) for the task and \(709\)--\(6042\) for local full state; these
are not end-to-end timing claims.

We compare against structured variants of these three adaptive methods, denoted
S-PAQT, S-SGQT, and S-OSGQT.  Each receives the same calibrated family, pilot,
and localization ball; S-PAQT retains Bayesian particle inference, while the others
return projected point iterates.  The primary experiment estimates the ordered Schmidt spectrum
across a \(3|3\) cut \cite{li_haldane_entanglement_spectrum}.  For every
dimension and copy budget, each adaptive method is granted its post hoc best
of four round counts, yet the fixed design
has lower mean task error in all 12 cells; 34 of the 36 paired intervals exclude
zero.  The accuracy advantage is not limited to the lower-dimensional task:
for full-state tomography, the fixed design also has lower mean infidelity than
every post hoc selected adaptive result in all eight calibrated local cells
\(R\le0.04\), with every paired interval excluding zero.  S-PAQT leads in
both families at the broadest radius, showing where fixed-centre linearization
ceases to be reliable.

The conceptual starting point is task-directed anchor geometry and
convex projection \cite{hay_sharon_next_manifolds,
hay_sharon_anchor_projection}.  We bring those ideas into quantum optimal
design, where Cram\'er--Rao criteria and convex optimization are established
tools \cite{kosut_walmsley_rabitz_oed,nunn_optimal_tomography}: the criterion
\(\Tr(Q\Sigma)\) is a task-weighted \(L\)-optimal analogue of
\(A\)-optimality \cite{pukelsheim_optimal_design}.  General approximate designs
may require randomized exchange \cite{harman_filova_richtarik_exchange}; here,
simultaneous diagonalization instead yields an exact integer allocation for the
retained probes.  This fixed-batch construction also differs from adaptive
Bayesian tomography, which selects each measurement through outcome-dependent
information gain \cite{huszar_houlsby_adaptive_bayesian}.  Classical shadows
lie outside the present access model because they use different measurements
\cite{aaronson_shadow_tomography,huang_kueng_preskill_classical_shadows}.

The main contributions are:
\begin{enumerate}
\item We derive the exact finite-shot Fisher information of physical fidelity
probes built from coordinate-invariant task directions, and bound the task loss
from omitting weak directions.
\item We propose Algorithm \ref{alg:spectral-fidelity-design}, a nonadaptive
spectral fidelity design with two proved optimality properties: each probe
location is the unique coordinatewise variance minimizer up to orientation,
and its greedy shot allocation is the exact global integer optimum over the
spectral library.  We also bound its gap to arbitrary-direction designs.
\item We give a closed-form offline estimator with a localization projection
that cannot increase local task loss when the assumed radius contains the
truth.
\item We demonstrate that our design attains lower local-regime error for both
task-directed and full-state tomography while requiring no feedback and
substantially lower acquisition cost.
\end{enumerate}

\section{Task Directions and Fidelity Probes}
\label{sec:local-task-tomography}

As described in the introduction, the available measurement primitive returns
a finite-shot estimate of the fidelity between an unknown pure state and a
chosen reference state.  The goal is to infer the state, or a specified
property of it, from a small collection of these scalar measurements.  Our
method constructs the reference states, called probes, whose fidelities reveal
the task-relevant directions, allocates the shot budget among them, and then
performs inference after the complete batch is collected.

This section develops the probe-construction step.  Around the reference state
\(\psi_0\), the matrix \(G\) measures how a parameter change moves the quantum
state, while \(Q\) measures how that change affects the requested output.  Their
generalized eigenvectors identify the directions that are most important for
the task, and we convert each such direction into a fidelity probe.  Section
\ref{sec:spectral-design} develops probe placement and shot allocation: it
proves that the optimal single probe for each task coordinate is aligned with
the corresponding generalized eigenvector, and then solves the integer shot
allocation over those probes exactly.

\subsection{State and task geometry}

We first distinguish changes in the state itself from changes that matter for
the requested output.  The two geometries below capture these notions and
identify the directions in which they differ most.

Let \(\Hcal\cong\C^D\), and suppose the unknown pure-state ray
\([\psi_\star]\in\mathbb P(\Hcal)\) belongs to a differentiable normalized
model
\[
\theta\longmapsto\psi(\theta),
\qquad
\theta\in\Theta\subseteq\R^d.
\]
Thus \([\psi_\star]=[\psi(\theta_\star)]\) for some unknown
\(\theta_\star\in\Theta\).
Global phase is immaterial, so all local geometry is defined on projective
Hilbert space.  Fix a design centre \(\theta_0\) and write
\(\psi_0=\psi(\theta_0)\).

For the parameter derivative \(\partial_j\psi_0\), define its horizontal
component
\[
D_j^\perp\psi_0
=
\left(I-\lvert\psi_0\rangle\langle\psi_0\rvert\right)
\partial_j\psi_0.
\]
The pullback Fubini--Study metric \cite{stokes_quantum_natural_gradient} is
\begin{equation}
G_{jk}=\operatorname{Re}\ip{D_j^\perp\psi_0}{D_k^\perp\psi_0},
\qquad
1-\abs{\ip{\psi(\theta_0)}{\psi(\theta_0+e)}}^2
=e^TGe+o(\norm{e}_2^2).
\label{eq:fs-pullback}
\end{equation}
We restrict to the parameter subspace on which \(G\succ0\), omitting directions
that do not change the state to first order.  The quadratic form \(e^TGe\)
therefore measures change in the state rather than arbitrary coordinate
displacement.

Let the requested scientific output be a differentiable, ray-invariant map
\[
A:\Theta\longrightarrow\R^m
\]
with a weight matrix \(W\succeq0\).  Its local squared error is represented by
\begin{equation}
Q=DA(\theta_0)^TWDA(\theta_0),
\qquad
\Lcal_Q(\hat\theta,\theta_\star)
=(\hat\theta-\theta_\star)^TQ(\hat\theta-\theta_\star).
\label{eq:task-metric}
\end{equation}
The formula is exact for a linear output and is the leading quadratic term for
a nonlinear output.  Full-state local infidelity is the special choice
\(Q=G\); in \(G\)-orthonormal coordinates it becomes \(Q=I\).

The important directions compare task sensitivity with the size of the
underlying state change.  Let
\begin{equation}
Qv_j=\lambda_jGv_j,
\qquad
v_i^TGv_j=\delta_{ij},
\qquad
\lambda_1\ge\cdots\ge\lambda_d\ge0.
\label{eq:generalized-task-spectrum}
\end{equation}
Equivalently, the symmetric matrix \(G^{-1/2}QG^{-1/2}\) has
ordinary orthonormal eigenvectors \(u_j=G^{1/2}v_j\).  Standard generalized
Rayleigh--Ritz theory \cite{golub_van_loan_matrix_computations} gives
\[
\lambda_j=\frac{v_j^TQv_j}{v_j^TGv_j}.
\]
Thus \(\lambda_j\) measures task sensitivity per unit of local state change in
direction \(v_j\).

The parameter coordinates are only a description of the state family, so
rescaling them or changing their basis should not alter the state
directions---and hence the probes---selected by the method.  The following
proposition makes this invariance precise.

\begin{proposition}[Local coordinate invariance]
\label{prop:coordinate-invariance}
Let \(\theta=\varphi(\eta)\) be a differentiable local reparameterization with
nonsingular Jacobian \(J_\varphi=D\varphi(\eta_0)\).  Then
\[
G_\eta=J_\varphi^TG_\theta J_\varphi,
\qquad
Q_\eta=J_\varphi^TQ_\theta J_\varphi.
\]
The generalized eigenvalues are unchanged, and if \(v_\theta\) is a
generalized eigenvector in the \(\theta\)-coordinates, then
\(v_\eta=J_\varphi^{-1}v_\theta\) represents the same tangent direction in
state space and has the same \(G\)-norm.
\end{proposition}

\begin{proof}
Both matrix transformations follow from the chain rule.  Substitution gives
\[
Q_\eta v_\eta
=J_\varphi^TQ_\theta v_\theta
=\lambda J_\varphi^TG_\theta v_\theta
=\lambda G_\eta v_\eta.
\]
Moreover,
\(v_\eta^TG_\eta v_\eta=v_\theta^TG_\theta v_\theta\), and the state-model
chain rule shows that their horizontal lifts coincide.  Hence the resulting
probe is independent of the chosen local chart.
\end{proof}

For an error \(e=\sum_jb_jv_j\), the two quadratic forms diagonalize as
\begin{equation}
e^TGe=\sum_{j=1}^db_j^2,
\qquad
e^TQe=\sum_{j=1}^d\lambda_jb_j^2.
\label{eq:task-coordinate-loss}
\end{equation}
Rapid decay in \(\lambda_j\) gives a low-effective-rank task even when the state
model itself is high dimensional.

\begin{definition}[Localization radius]
\label{def:localization-radius}
For a retained displacement \(e\), define
\[
r_G(e):=\sqrt{e^TGe},
\qquad
\mathcal B_R(\theta_0)
:=\{\theta_0+e:r_G(e)\le R\}
=\left\{\theta_0+\sum_jb_jv_j:\norm{b}_2\le R\right\}.
\]
Thus \(R\) is a dimensionless local Fubini--Study bound, not a Euclidean radius
in the original Hamiltonian parameters.  For a supplied pre-data admissible set
\(\Theta_{\rm loc}\), calculate
\(R=\sup_{\theta\in\Theta_{\rm loc}}r_G(\theta-\theta_0)\); alternatively, a
target boundary infidelity \(\varepsilon\) gives
\(R\simeq\sqrt{\varepsilon}\) through \eqref{eq:fs-pullback}.  It is not
estimated from the subsequent outcomes.
\end{definition}

\begin{proposition}[Loss from omitted task directions]
\label{prop:task-tail}
Suppose \(e=\sum_jb_jv_j\), \(e^TGe\le \Delta^2\), and the leading coordinates
satisfy \(\sum_{j=1}^rb_j^2\le\eta^2\).  Then
\[
\Lcal_Q(\hat\theta,\theta_\star)
\le
\lambda_{r+1}\Delta^2
+(\lambda_1-\lambda_{r+1})\min\{\eta^2,\Delta^2\},
\]
where \(\lambda_{d+1}=0\) if \(r=d\).  More precisely,
\[
\Lcal_Q(\hat\theta,\theta_\star)
\le
\lambda_{r+1}\Delta^2
+\sum_{j=1}^r(\lambda_j-\lambda_{r+1})b_j^2.
\]
\end{proposition}

\begin{proof}
Split the second identity in \eqref{eq:task-coordinate-loss} at \(r\) and use
\(\lambda_j\le\lambda_{r+1}\) for \(j>r\).  Adding and subtracting
\(\lambda_{r+1}\sum_{j\le r}b_j^2\) gives the sharper bound.  The scalar bound
then follows from \(\sum_{j\le r}b_j^2\le\min\{\eta^2,\Delta^2\}\) and
\(\lambda_j-\lambda_{r+1}\le\lambda_1-\lambda_{r+1}\).
\end{proof}

The proposition explains the intended task advantage.  When
\(\lambda_{r+1}\) is small, measurements that control the first \(r\) task
coordinates suppress most of the loss without uniformly identifying the
remaining state directions.  Full-state tomography instead includes every
identifiable state direction.

This focus on the leading task directions is the local quantum analogue of the
output-directed anchor constructions in \cite{hay_sharon_next_manifolds,
hay_sharon_anchor_projection}, where anchors are chosen for their relevance to
a downstream quantity rather than for uniform reconstruction.

\subsection{From task directions to fidelity probes}

The important directions identified above are initially expressed in parameter
space.  To measure them experimentally, we now turn each direction into a probe
state whose fidelity changes with motion along that direction.

Each probe is a known reference state whose fidelity with the unknown state is
measured.  Such reference states are called anchor states in the distance-based
frameworks of \cite{hay_sharon_next_manifolds,
hay_sharon_anchor_projection}.

For a \(G\)-unit direction \(w=\sum_jc_jv_j\), meaning \(w^TGw=1\), define its
horizontal lift
\[
u_w=\sum_k w_kD_k^\perp\psi_0.
\]
Then \(u_w\perp\psi_0\) and \(\norm{u_w}=1\).  For a radius
\(0<\rho<\pi/2\), the corresponding geodesic gives the probe
\begin{equation}
\phi(\rho,w)=\cos\rho\,\psi_0+\sin\rho\,u_w.
\label{eq:geodesic-probe}
\end{equation}
Its fidelity with \(\psi_0\) is
\(\abs{\ip{\phi(\rho,w)}{\psi_0}}^2=\cos^2\rho\), and its Fubini--Study
distance from \([\psi_0]\) is \(\rho\).  Replacing \(w\) by \(-w\) reverses
the orientation.

The probe need not lie in the parametric family of target states; it is a
reference state used for the fidelity measurement.

Write a nearby state as
\[
\theta(b)=\theta_0+\sum_jb_jv_j,
\qquad
F_w(b)=\abs{\ip{\phi(\rho,w)}{\psi(\theta(b))}}^2.
\]
Here \(b=(b_1,\ldots,b_d)^T\) is the local displacement from \(\theta_0\) in
the generalized eigenvector basis; \(b_j\) is the signed displacement along
\(v_j\).
Differentiation at \(b=0\) yields
\begin{equation}
F_w(0)=\cos^2\rho,
\qquad
\nabla_bF_w(0)=\sin(2\rho)c.
\label{eq:geodesic-score}
\end{equation}
The nonzero slope makes the probe informative about signed displacement along
\(w\); its distance from \(\psi_0\) alone would not distinguish the two
orientations.

\section{Spectral Experimental Design}
\label{sec:spectral-design}

Section \ref{sec:local-task-tomography} constructs the eigenvector probes but
does not specify an acquisition plan.  We now decide which probes to use and
how many shots to assign to each one.  We first state the Bayesian design
objective, then derive the local criterion used by the algorithm and solve the
resulting integer allocation problem.

\subsection{Design objective and local approximation}

The design goal is to allocate a shot budget so that the posterior is as
concentrated as possible in the directions that matter for the requested
output.  We first specify the data generated by a candidate batch and then
formulate the objective used to compare candidate designs.

\begin{definition}[Fidelity batch design and observation model]
\label{def:fidelity-batch-observation-model}
Let \(\mathcal W=\{w\in\R^d:w^TGw=1\}\), and fix a radius \(\rho\) and shot
budget \(B\).  A batch design consists of \(K\ge1\) directions and their integer
shot counts,
\[
\mathbf w=(w_1,\ldots,w_K)\in\mathcal W^K,
\qquad
N=(N_1,\ldots,N_K)\in\mathbb Z_{\ge0}^K,
\qquad
\sum_{i=1}^KN_i=B.
\]
At each probe \(\phi_i=\phi(\rho,w_i)\), the observed success count is
\begin{equation}
X_i\mid\theta
\sim
\operatorname{Binomial}(N_i,p_i(\theta)),
\qquad
p_i(\theta)=h_i(F_i(\psi(\theta))),
\label{eq:binomial-model}
\end{equation}
where the known calibration map \(h_i\) converts fidelity to the observed
success probability.  Direct projective readout has \(h_i(F)=F\); a swap test
has \(h_i(F)=(1+F)/2\)
\cite{nielsen_chuang_qcqi,buhrman_quantum_fingerprinting}.  We assume that the
counts are conditionally independent given \(\theta\).  Writing
\(Z=(X_1,\ldots,X_K)\), their joint likelihood is
\begin{equation}
L^{\rm fid}_{\mathbf w,N}(\theta;Z)
=
\prod_{i=1}^{K}
\binom{N_i}{X_i}
p_i(\theta)^{X_i}
\{1-p_i(\theta)\}^{N_i-X_i}.
\label{eq:fidelity-likelihood}
\end{equation}
\end{definition}

The conditional-independence assumption excludes shared calibration uncertainty
or drift, which would require a joint observation model.  This likelihood allows
a proposed batch design to be evaluated before data are collected by averaging
its posterior uncertainty over all possible outcomes.

\begin{definition}[Bayesian fidelity-design problem]
\label{def:bayesian-fidelity-design-problem}
Fix a prior \(p_0\) and a measure \(D\) of posterior spread.  For a batch design
\((\mathbf w,N)\) and observed counts \(Z\), define the posterior by
\[
\pi_{\mathbf w,N}(d\theta\mid Z)
\propto
p_0(\theta)L^{\rm fid}_{\mathbf w,N}(\theta;Z)\,d\theta.
\]
The Bayesian fidelity-design problem is
\[
\inf_{\substack{K\ge1,\;\mathbf w\in\mathcal W^K\\
                 N\in\mathbb Z_{\ge0}^K,\;\sum_iN_i=B}}
\mathbb E_{Z\mid\mathbf w,N}
\left[
 D\{\pi_{\mathbf w,N}(\cdot\mid Z)\}
\right].
\]
The expectation is taken under the prior-predictive distribution determined by
\(p_0\) and the finite-shot likelihood.
\end{definition}

Definition~\ref{def:bayesian-fidelity-design-problem} is a standard Bayesian
experimental-design objective \cite{chaloner_verdinelli_bayesian_design}.

To direct this objective toward the requested output, we now choose the posterior
spread measure \(D\) using the task matrix \(Q\).

\begin{definition}[Task-weighted posterior spread]
\label{def:task-posterior-spread}
Let \(\pi\) be a posterior with finite second moments, mean
\(\bar\theta_\pi\), and covariance
\[
\Sigma_\pi
=
\mathbb E_\pi\!\left[
(\theta-\bar\theta_\pi)(\theta-\bar\theta_\pi)^T
\right].
\]
Given the task matrix \(Q\succeq0\), define the task-weighted posterior spread
by
\[
\Phi_Q(\pi)
=
\Tr(Q\Sigma_\pi)
=
\mathbb E_\pi\!\left[
(\theta-\bar\theta_\pi)^TQ(\theta-\bar\theta_\pi)
\right].
\]
Posterior \(\pi_2\) has no greater task-weighted spread than \(\pi_1\) when
\(\Phi_Q(\pi_2)\le\Phi_Q(\pi_1)\).
\end{definition}

With \(D=\Phi_Q\), Definition~
\ref{def:bayesian-fidelity-design-problem} becomes the global Bayes-risk
criterion
\begin{equation}
\mathfrak R_Q(\mathbf w,N)
:=
\mathbb E_{Z\mid\mathbf w,N}
\left[
\Phi_Q\{\pi_{\mathbf w,N}(\cdot\mid Z)\}
\right].
\label{eq:global-bayes-design-risk}
\end{equation}
For the posterior-mean estimator, \(\mathfrak R_Q\) is the task-weighted
estimation error expected from a proposed design.  It is the ideal objective
that the local criterion below approximates.

Direct evaluation of \eqref{eq:global-bayes-design-risk} may be costly because
it averages an outcome-dependent posterior covariance over the prior-predictive
distribution.  We therefore use a local covariance approximation based on
Fisher information.  Let
\(\Lambda_0\succ0\) be a prior-precision matrix at the design centre.  For a
smooth prior, this is the Hessian of its negative log density; for a Gaussian
prior it is the inverse covariance.  For the bounded particle prior used in the
experiments below, we instead use the inverse prior covariance.  The resulting
covariance approximation is
\[
\Sigma_{\rm loc}(\mathbf w,N)
=
\left(
\Lambda_0+\sum_i\mathcal I_i(\theta_0)
\right)^{-1},
\]
where \(\mathcal I_i(\theta_0)\) is the information contributed by the
\(N_i\) shots at probe \(i\).  Substituting this approximation into the task
loss gives
\begin{equation}
\mathfrak R_Q(\mathbf w,N)
\approx
\Psi_Q^{\rm loc}(\mathbf w,N)
:=
\Tr\{Q\Sigma_{\rm loc}(\mathbf w,N)\}.
\label{eq:local-task-objective}
\end{equation}
Thus the observation likelihood enters the local design through the information
matrices \(\mathcal I_i(\theta_0)\).  Once those matrices are derived, the
remaining optimization depends on the likelihood only through this local
curvature.
This local covariance approximation is exact in the corresponding
linear-Gaussian model.  In the nonlinear setting, it is reliable only when the
prior-predictive mass remains in a region with stable likelihood curvature and
no unresolved posterior modes.

The global risk can instead be evaluated by nested or simulation-based
Bayesian design \cite{huan_marzouk_simulation_design}, but that calculation is
substantially more expensive and is not the acquisition algorithm studied
here.

\subsection{Finite-shot information and spectral probes}

Having fixed the local design criterion, we now compute how much information a
finite number of measurements provides for each probe.  A fidelity measurement
produces a binary outcome, so its usefulness depends on how sensitively its
success probability changes with the unknown state.

Recall that \(b_j\) is the signed displacement from the design centre along
the generalized direction \(v_j\).  For \(w=\sum_jc_jv_j\), the vector \(c\)
specifies the probe direction in these local coordinates.  The next proposition
shows that one shot contributes \(\nu_\rho\) units of information in this
direction and no first-order information in directions orthogonal to it.

\begin{proposition}[Information of a geodesic probe]
\label{prop:fidelity-information}
Let \(w=\sum_jc_jv_j\), and collect \(N\) shots from the probe
\(\phi(\rho,w)\) using the binomial model \eqref{eq:binomial-model}.  Its
Fisher information in the local displacement coordinates \(b\), evaluated at
the design centre \(b=0\), is
\[
\mathcal I_b^{\rm fid}(w;0)=N\nu_\rho cc^T,
\qquad
\nu_\rho=
\frac{h'(\cos^2\rho)^2\sin^2(2\rho)}
{h(\cos^2\rho)\{1-h(\cos^2\rho)\}}.
\]
For direct projective readout, \(\nu_\rho=4\) for every
\(0<\rho<\pi/2\).  For the swap test,
\(\nu_\rho=4\cos^2\rho/(1+\cos^2\rho)\).
\end{proposition}

\begin{proof}
Write \(p=p(b)=h(F_w(b))\), and let \(X\) be the observed success count.  The
binomial log-likelihood and its score are
\[
\ell(b)=X\log p+(N-X)\log(1-p),
\qquad
\nabla_b\ell(b)=\frac{X-Np}{p(1-p)}\nabla_bp.
\]
Since \(\mathbb E[(X-Np)^2]=Np(1-p)\), taking the expected outer product of
the score gives
\[
\mathcal I_b(b)
=\mathbb E\!\left[\nabla_b\ell(b)\nabla_b\ell(b)^T\right]
=\frac{N}{p(b)\{1-p(b)\}}\nabla_bp(b)\nabla_bp(b)^T.
\]
At the design centre, \eqref{eq:geodesic-score} and the chain rule give
\[
p(0)=h(\cos^2\rho),
\qquad
\nabla_bp(0)=h'(\cos^2\rho)\sin(2\rho)c.
\]
Substitution yields
\[
\mathcal I_b^{\rm fid}(w;0)
=N
\frac{h'(\cos^2\rho)^2\sin^2(2\rho)}
{h(\cos^2\rho)\{1-h(\cos^2\rho)\}}
cc^T
=N\nu_\rho cc^T.
\]
For direct projective readout, \(h(F)=F\), and therefore
\[
\nu_\rho
=\frac{\sin^2(2\rho)}{\cos^2\rho\sin^2\rho}
=4.
\]
For the swap test, \(h(F)=(1+F)/2\), and direct simplification gives
\[
\nu_\rho
=\frac{\tfrac14\sin^2(2\rho)}
{\tfrac12(1+\cos^2\rho)\tfrac12\sin^2\rho}
=\frac{4\cos^2\rho}{1+\cos^2\rho}.
\]
\end{proof}

\begin{remark}[Why the experiments use \(\rho=\pi/4\)]
\label{rem:pi-over-four}
For direct projective readout, \(\nu_\rho=4\) means that the leading Fisher
term alone does not select a radius.  We use \(\rho=\pi/4\) as a
well-conditioned finite-sample choice.  If the unknown state equals the design
centre, then a direct comparison with the probe succeeds with probability
\(F_w(0)=\cos^2(\pi/4)=1/2\).  Thus repeated shots are balanced between
successes and failures.  A small displacement toward the probe raises this
probability above \(1/2\), while a displacement away lowers it, and the local
slope \(\sin(2\rho)=1\) is maximal.

The experiment observes a success count \(X\), and hence the frequency
\(\widehat p=X/N\), rather than the signed displacement \(z=c^Tb\) directly.
Along the local geodesic relation, \(p=\cos^2(\rho-z)\); solving for \(z\)
defines
\[
z=g_\rho(p):=\rho-\arccos\sqrt p.
\]
Thus \(g_\rho\) is simply the calibration that converts an observed success
probability back into displacement along the probe direction.  Passing the
noisy frequency \(\widehat p\) through a curved calibration can introduce
finite-sample bias.  At \(\rho=\pi/4\),
\(g_{\pi/4}''(1/2)=0\), so the calibration is locally linear and its leading
bias term vanishes.

\end{remark}

The design algorithm makes two choices: which directions to probe and how to
divide the shot budget among them.  Proposition
\ref{prop:fidelity-information} shows that \(N\) shots in direction
\(w=\sum_kc_kv_k\) add the rank-one information \(N\nu_\rho cc^T\).  The next
theorem uses this information to determine which single probe direction gives
the smallest local variance for a chosen coordinate \(b_j\).  It allows the
prior certainty to differ across coordinates, provided the prior precision is
diagonal in this basis.

\pagebreak
\begin{theorem}[Coordinatewise optimality of spectral probes]
\label{thm:one-probe-directional-optimality}
Fix a task coordinate \(b_j\), a radius \(0<\rho<\pi/2\), and a shot count
\(N\ge1\).  For any \(G\)-unit direction \(w=\sum_kc_kv_k\), let
\(c=(c_1,\ldots,c_d)^T\), so \(c^Tc=1\).  Assume \(\nu_\rho>0\), set
\(\tau:=N\nu_\rho\), and take the baseline local precision in the
\(b\)-coordinates to be
\(\Lambda_0=\operatorname{diag}(\alpha_1,\ldots,\alpha_d)\), with
\(\alpha_k>0\).  The local covariance produced by measuring the probe
\(\phi(\rho,w)\) for \(N\) shots is
\[
\Sigma_{\rm loc}(w)
:=
(\Lambda_0+\tau cc^T)^{-1}.
\]
Then the local variance of coordinate \(b_j\) is
\[
V_j(w)
:=
e_j^T\Sigma_{\rm loc}(w)e_j
=
\frac1{\alpha_j}
-
\frac{\tau c_j^2/\alpha_j^2}
{1+\tau\sum_kc_k^2/\alpha_k}.
\]
Consequently,
\[
V_j(w)\ge V_j(v_j)=\frac1{\alpha_j+\tau},
\]
with equality if and only if \(w=\pm v_j\).
\end{theorem}

\begin{proof}
By Proposition~\ref{prop:fidelity-information}, the \(N\) shots
contribute Fisher information \(\tau cc^T\).  Adding this to the baseline
precision \(\Lambda_0\) gives the precision matrix in the definition of
\(\Sigma_{\rm loc}(w)\).

The Sherman--Morrison formula gives
\[
\Sigma_{\rm loc}(w)
=
\Lambda_0^{-1}
-
\frac{\tau\Lambda_0^{-1}cc^T\Lambda_0^{-1}}
{1+\tau c^T\Lambda_0^{-1}c}.
\]
Taking the \(j\)th diagonal entry yields
\[
V_j(w)
=
\frac1{\alpha_j}
-
\frac{\tau c_j^2/\alpha_j^2}
{1+\tau\sum_kc_k^2/\alpha_k}.
\]
The variance reduction is largest at \(c=\pm e_j\), because
\[
\frac{\tau c_j^2/\alpha_j^2}
{1+\tau\sum_kc_k^2/\alpha_k}
\le
\frac{\tau}{\alpha_j(\alpha_j+\tau)}.
\]
After clearing the positive denominators, this inequality is equivalent to
\[
\alpha_j(1-c_j^2)
+
\alpha_j\tau\sum_{k\ne j}\frac{c_k^2}{\alpha_k}
\ge0.
\]
Equality holds if and only if \(c_k=0\) for every \(k\ne j\), hence
\(c=\pm e_j\), or equivalently \(w=\pm v_j\).  Substitution gives
\(V_j(v_j)=1/(\alpha_j+\tau)\).
\end{proof}

This theorem explains the choice of spectral probes one coordinate at a
time: for a fixed radius and shot count, the probe aligned with \(v_j\) is the
best single setting for reducing uncertainty in \(b_j\).  It does not yet solve
the joint task problem, because a probe aligned with a combination of several
\(v_j\) may reduce several coordinates simultaneously.  Theorem
\ref{thm:spectral_cost} later quantifies the large-budget loss from
excluding such directions.

\subsection{Exact integer allocation over the spectral probes}
\label{subsec:exact-spectral-allocation}

The preceding results motivate using the first \(r\) eigenvector probes
\(v_1,\ldots,v_r\).  We now fix this set and optimize their shot counts jointly.
The following posterior decomposition explains why the task eigenvalues become
the weights in this restricted allocation problem.

\begin{corollary}[Posterior task-loss decomposition]
\label{cor:posterior-task-tail}
Let \(\pi\) have finite second moments and posterior mean
\(\bar\theta_\pi\).  Define the centered task coordinate and its posterior
variance by
\[
\widetilde b_j(\theta)
=
v_j^TG(\theta-\bar\theta_\pi),
\qquad
V_j
=
\mathbb E_\pi[\widetilde b_j(\theta)^2].
\]
Then
\begin{equation}
\Phi_Q(\pi)
=
\sum_{j=1}^d\lambda_jV_j.
\label{eq:task-variance-decomposition}
\end{equation}
For \(1\le r<d\), if \(\sum_{j>r}V_j\le \tau_{\rm tail}^2\), then
\[
\Phi_Q(\pi)
\le
\sum_{j=1}^r\lambda_jV_j
+
\lambda_{r+1}\tau_{\rm tail}^2.
\]
\end{corollary}

\begin{proof}
Because the generalized task basis is \(G\)-orthonormal,
\(\theta-\bar\theta_\pi=\sum_j\widetilde b_j(\theta)v_j\), and the quadratic
task loss is \(\sum_j\lambda_j\widetilde b_j(\theta)^2\).  Taking posterior
expectation proves \eqref{eq:task-variance-decomposition}.  For the tail bound,
use \(\lambda_j\le\lambda_{r+1}\) for \(j>r\).
\end{proof}

Definition \ref{def:bayesian-fidelity-design-problem} allows the probe
directions and their shot counts to vary and evaluates the resulting global
Bayesian risk.  We now restrict that problem by fixing the candidate probes to
\(v_1,\ldots,v_r\) and replacing the global risk by the local covariance
criterion \eqref{eq:local-task-objective}.  When the prior precision is diagonal
in this basis and each retained probe has known per-shot information, only the
integer shot counts remain to be chosen.

\begin{definition}[Spectral shot-allocation problem]
\label{def:spectral-shot-allocation}
Fix \(1\le r\le d\), prior precision
\(\Lambda_0=\operatorname{diag}(\alpha_1,\ldots,\alpha_d)\) with
\(\alpha_j>0\), per-shot information \(\nu_j>0\) for probe \(v_j\), and an
integer budget \(B\in\mathbb Z_{\ge0}\).  For
\(n=(n_1,\ldots,n_r)\in\mathbb Z_{\ge0}^r\), where \(n_j\) is the number of
shots assigned to probe \(v_j\), the prior and measurement precisions add, so
the local variance in direction \(v_j\) is
\((\alpha_j+\nu_j n_j)^{-1}\).  Corollary
\ref{cor:posterior-task-tail} therefore gives the local task-variance surrogate
\begin{equation}
\Psi_r(n)
=
\sum_{j=1}^r\frac{\lambda_j}{\alpha_j+\nu_j n_j}
+
\sum_{j=r+1}^d\frac{\lambda_j}{\alpha_j}.
\label{eq:spectral-risk}
\end{equation}
The first sum is the task-weighted local variance of the retained directions
after allocation, while the second is the prior-only contribution from the
excluded directions \(j>r\).  The spectral shot-allocation problem is to choose
\[
n^\star
\in
\argmin_{\substack{n\in\mathbb Z_{\ge0}^r\\\sum_{j=1}^r n_j=B}}
\Psi_r(n).
\]
Thus \(n^\star\) denotes any allocation attaining the minimum.
\end{definition}

A common radius makes \(\nu_j\) direction-independent for both readout models
(Proposition \ref{prop:fidelity-information}); for direct readout, \(\nu_j=4\)
for any radius.  An isotropic local prior further gives \(\alpha_j=\alpha\).  A
fixed ordering by \(\lambda_j\) is not sufficient once shots have been assigned:
a direction with a large task weight should be measured early, but its priority
decreases as its variance is reduced.

\begin{proposition}[Marginal task-variance reduction]
\label{prop:marginal-task-variance-reduction}
Let \(\alpha_j>0\), \(\nu_j>0\), \(\lambda_j\ge0\), and
\(k\in\mathbb Z_{\ge0}\).  The decrease in the task-variance contribution
from assigning the next shot to direction \(v_j\) is
\begin{equation}
\Delta_j(k)
=
\frac{\lambda_j}{\alpha_j+\nu_j k}
-
\frac{\lambda_j}{\alpha_j+\nu_j(k+1)}
=
\frac{\lambda_j\nu_j}
{(\alpha_j+\nu_j k)(\alpha_j+\nu_j(k+1))}.
\label{eq:marginal-gain}
\end{equation}
Moreover,
\[
0\le\Delta_j(k+1)\le\Delta_j(k).
\]
\end{proposition}

\begin{proof}
Taking a common denominator gives the last expression in
\eqref{eq:marginal-gain}.  Its numerator is nonnegative, while the positive
denominator increases with \(k\).  Thus the marginal reductions are
nonnegative and nonincreasing.
\end{proof}

\begin{remark}[Effect of prior and measurement precision]
\label{rem:isotropic-prior-precision-allocation}
The prior supplies precision \(\alpha_j\) in direction \(v_j\), while the
\(n_j\) assigned shots add precision \(\nu_jn_j\).  Thus a direction receives
lower priority when it is already well known, and higher priority when its
probe is especially informative.  In the common case
\(\alpha_j=\alpha\) and \(\nu_j=\nu\), the ratio
\begin{equation}
\Gamma:=\frac{\nu B}{\alpha}
\label{eq:information-to-prior-ratio}
\end{equation}
compares the precision available from the entire batch with the prior precision
in one direction.  If \(\Gamma\ll1\), even concentrating the full batch in one
direction changes its total precision only slightly, so the optimal allocation
tends to remain concentrated on the largest task weights.  If \(\Gamma\gg1\),
measured directions become precise enough that their marginal gains fall below
those of weaker directions, and the allocation spreads.
\end{remark}

\begin{remark}[Choosing the number of task directions]
The number \(r\) is a scientific choice, not an output of the allocation proof.
One may use all directions with positive task eigenvalues, limit the number of
probes before acquisition, or choose the smallest \(r\) for which the final sum
in \eqref{eq:spectral-risk} is below a specified task-risk tolerance.  The
contribution \(\sum_{j>r}\lambda_j/\alpha_j\) remains explicit.  Full-state
tomography is the case \(Q=G\): then \(\lambda_j=1\), \(r=d_{\rm loc}\), and
\eqref{eq:marginal-gain} gives a balanced allocation over all identifiable
directions.
\end{remark}

Theorem \ref{thm:one-probe-directional-optimality} justifies the eigenvector
probes used in Step 2 coordinate by coordinate, while Proposition
\ref{prop:marginal-task-variance-reduction} gives the marginal reductions used
in Step 3.  Together, they give a simple rule: construct the first \(r\)
eigenvector probes and assign each new shot to the probe with the largest
current reduction \(\Delta_j\).  Theorem \ref{thm:greedy-optimality} below
certifies the resulting allocation.  All probes and shot counts are chosen
before any outcomes are observed.

\begin{algorithm}[H]
\caption{Task-directed spectral fidelity design}
\label{alg:spectral-fidelity-design}
\textbf{Input.} State model \(\psi(\theta)\); design centre \(\theta_0\); task
map \(A\) and weight \(W\); number of directions \(r\); radius \(\rho\); prior
precisions \(\alpha_1,\ldots,\alpha_d\) in the task coordinates; per-shot
information \(\nu_1,\ldots,\nu_r\); shot budget \(B\).
\begin{enumerate}
\item Compute the pullback metric \(G\), task metric \(Q\), and the
\(G\)-orthonormal generalized eigenpairs in
\eqref{eq:generalized-task-spectrum}.
\item For \(j=1,\ldots,r\), construct the probe
\(\phi_j=\phi(\rho,v_j)\) from \eqref{eq:geodesic-probe}.
\item Initialize \(n_j=0\).  For \(t=1,\ldots,B\), choose the smallest index
in
\[
\argmax_{1\le j\le r}\Delta_j(n_j)
\]
and increment its allocation.
\end{enumerate}
\textbf{Output.} The fixed batch design
\(\mathcal D^\star=\{(\phi_j,n_j):1\le j\le r,\ n_j>0\}\).
\end{algorithm}
\vspace{-\baselineskip}
\begin{theorem}[Exact spectral allocation]
\label{thm:greedy-optimality}
In the setting of Definition \ref{def:spectral-shot-allocation}, the count
vector \(n_j\) produced in Step 3 of Algorithm
\ref{alg:spectral-fidelity-design} is a global minimizer of the spectral
shot-allocation problem.
\end{theorem}

\begin{proof}
For any allocation satisfying the constraints, telescoping
\eqref{eq:marginal-gain} gives
\[
\sum_{j=1}^r\frac{\lambda_j}{\alpha_j}
-
\sum_{j=1}^r\frac{\lambda_j}{\alpha_j+\nu_j n_j}
=
\sum_{j=1}^r\sum_{k=0}^{n_j-1}\Delta_j(k).
\]
Thus an allocation selects a prefix of length \(n_j\) from each decreasing
marginal-gain sequence.  The tail term
\(\sum_{j=r+1}^d\lambda_j/\alpha_j\) in \eqref{eq:spectral-risk} is constant, so
minimizing \(\Psi_r\) is equivalent to maximizing the telescoping reduction.
At any intermediate allocation, every unselected gain in sequence \(j\) is no
larger than its current head \(\Delta_j(n_j)\).  Step 3 therefore selects a
largest remaining gain at each iteration.  After \(B\) iterations,
\(n^{\rm gr}\) maximizes the total reduction and hence globally minimizes
\(\Psi_r\).
\end{proof}

Because \(\Psi_r\) depends only on the final shot counts, outcome feedback cannot
improve this fixed-centre surrogate over the same probe library.  This
observation does not extend to the global Bayesian risk
\eqref{eq:global-bayes-design-risk}, where outcomes may change the posterior
geometry or motivate recentering.

\begin{remark}[Implementation and optimality certificate]
A max heap keyed by \(\Delta_j(n_j)\) implements the rule in
\(O(r+B\log r)\) time and \(O(r)\) memory after the eigendecomposition.  This
is typically small relative to state-derivative evaluation or posterior
inference.  The allocation can also be characterized by a marginal threshold:
at the optimum, every selected marginal gain is no smaller than any unselected
one.  This gives a direct certificate for the computed allocation and is
convenient for regression tests.
\end{remark}

\begin{remark}[Scope of exact spectral optimality]
The exact greedy result allows unequal prior precisions \(\alpha_j\) and
per-shot information coefficients \(\nu_j\).  It requires the prior and
measurement information to be diagonal in the same task basis.  Off-diagonal
prior precision, correlated shot outcomes, or information directions that vary
with the unknown state instead produce a coupled matrix-design problem and
require optimizing \eqref{eq:local-task-objective} directly.

The exact result is restricted to eigenvector probes; the next subsection shows
that the finite-budget advantage of arbitrary directions vanishes as the shot
count grows.
\end{remark}

\subsection{Continuous relaxation and the cost of the spectral restriction}

The finite-budget gap appears already in two dimensions.  In \(G\)-orthonormal
coordinates, let
\(\widetilde Q=\operatorname{diag}(4,1)\),
\(\alpha_1=\alpha_2=\nu=1\), and \(B=2\).  The best eigenvector-only value is
\(7/3\), whereas the two mixed directions
\(w_\pm=(\sqrt{5/6},\,\pm\sqrt{1/6})^T\) give \(9/4\).  This improvement uses
two distinct mixed probes, while the best eigenvector design repeats a single
probe.

This example motivates the question addressed by Theorem~\ref{thm:spectral_cost}: does the advantage of mixed directions remain substantial as the shot count grows?  To compare the two probe classes, retain the diagonal prior precision
\(\Lambda_0=\operatorname{diag}(\alpha_1,\ldots,\alpha_d)\), but assume a common
per-shot information coefficient \(\nu_j=\nu\).  Proposition
\ref{prop:fidelity-information} shows that this assumption holds for both direct
projective readout and the swap test when every direction uses the same probe
radius.  The common coefficient assigns information \(\nu ww^T\) to any unit
probe direction \(w\), including a mixture of task eigenvectors.  Suppose the
positive task eigenvalues are
\(\lambda_1\ge\cdots\ge\lambda_r>0\), and define
\begin{align}
\Psi_{\rm all}(B)
&:=
\min_{\substack{w_1,\ldots,w_B\in\R^d\\ \norm{w_t}_2=1}}
\Tr\!\left[
\widetilde Q
\left(\Lambda_0+\nu\sum_{t=1}^B w_tw_t^T\right)^{-1}
\right],
\label{eq:unrestricted-shot-optimum}\\
\Psi_{\rm spec}(B)
&:=
\min_{\substack{n_j\in\mathbb Z_{\ge0}\\\sum_{j=1}^r n_j=B}}
\sum_{j=1}^r\frac{\lambda_j}{\alpha_j+\nu n_j}.
\label{eq:spectral-shot-optimum}
\end{align}
Here \(\widetilde Q=\operatorname{diag}(\lambda_1,\ldots,\lambda_r,0,\ldots,0)\).
The first objective permits an arbitrary unit direction on every shot; the
second reuses at most \(r\) eigenvector probes.

Define the fractional spectral relaxation
\begin{equation}\label{eq:fractional_spectral}
\Psi_{\rm frac}(B)
:=
\min_{\substack{s\in\mathbb{R}_+^r\\ \mathbf{1}^{T}s=B}}
\sum_{j=1}^r \frac{\lambda_j}{\alpha_j+\nu s_j}.
\end{equation}
Its minimizer is the active-set water-filling solution
$s_j^\star=(c\sqrt{\lambda_j}-\alpha_j)_+/\nu$, where $c$ is chosen so that $\sum_j s_j^\star=B$. Set
$$ L:=\sum_{j=1}^r\sqrt{\lambda_j}, \quad \alpha_\Sigma:=\sum_{j=1}^r\alpha_j, \quad 
 \text{and} \quad c_B:=(\nu B+\alpha_\Sigma)/L . $$
If all coordinates are active, then $c=c_B$ and
$\Psi_{\rm frac}(B)=L^2/(\nu B+\alpha_\Sigma)$.



\begin{theorem}[Finite- and large-budget cost of spectral integer allocation]
\label{thm:spectral_cost}
For every integer budget $B$,
\begin{equation}\label{eq:fractional_certificate}
\Psi_{\rm frac}(B)
\leq
\Psi_{\rm all}(B)
\leq
\Psi_{\rm spec}(B).
\end{equation}
If
\[
c_B\sqrt{\lambda_j}\geq \max\{\alpha_j+\nu,2\nu\},
\qquad j=1,\ldots,r,
\]
then denote $H:=\sum_{j=1}^r\lambda_j^{-1/2}$ and obtain
\begin{align*}
0\le
\Psi_{\rm spec}(B)-\Psi_{\rm all}(B)
&\le
\frac{2\nu^2L^3H}{(\nu B+\alpha_\Sigma)^3},\\
1\le
\frac{\Psi_{\rm spec}(B)}{\Psi_{\rm all}(B)}
&\le
1+\frac{2\nu^2LH}{(\nu B+\alpha_\Sigma)^2}.
\end{align*}
Thus, the absolute and relative gaps are respectively \(O(B^{-3})\) and \(O(B^{-2})\) for fixed spectrum, prior precisions, and per-shot information.
\end{theorem}

\begin{proof}
Let \(w_1^\star,\ldots,w_B^\star\) attain \(\Psi_{\rm all}(B)\), let \(e_j\)
be the coordinate vector for eigenvector \(v_j\), and set
\[
A=\Lambda_0+\nu\sum_{t=1}^B w_t^\star(w_t^\star)^T.
\]
Writing each \(w_t^\star\) in the eigenvector basis, its aggregate fractional
contribution to direction \(j\) is
\[
s_j:=\sum_{t=1}^B(e_j^Tw_t^\star)^2,
\qquad
\sum_{j=1}^d s_j=B.
\]
The diagonal precision in direction \(j\) is therefore
\(e_j^TAe_j=\alpha_j+\nu s_j\).  Cauchy--Schwarz gives
\[
(e_j^TAe_j)(e_j^TA^{-1}e_j)\ge1,
\qquad
e_j^TA^{-1}e_j\ge\frac{1}{\alpha_j+\nu s_j}.
\]
Thus deleting the cross terms of the optimal mixed design gives a fractional
eigenvector allocation with no larger task risk.  Since any contribution to
directions \(j>r\) can instead be assigned to a positive-weight direction,
\[
\Psi_{\rm all}(B)
\geq
\sum_{j=1}^r\frac{\lambda_j}{\alpha_j+\nu s_j}
\geq
\Psi_{\rm frac}(B).
\]
Since every spectral design is admissible in the arbitrary-direction problem, $\Psi_{\rm all}(B)\leq\Psi_{\rm spec}(B)$, proving~\eqref{eq:fractional_certificate}.

For the quantitative bounds, assume the stated condition. Then all
coordinates are active,
$\Psi_{\rm frac}(B)=L^2/(\nu B+\alpha_\Sigma)$, and every
$s_j^\star\geq1$.
We now replace the fractional counts with integer counts. Since \(s_j^*\ge1\), we round them to nonnegative integers
\(n_j\) with \(\sum_jn_j=B\) and \(\lvert n_j-s_j^*\rvert<1\), and write
\(\delta_j=n_j-s_j^*\).  Then, \(\sum_j\delta_j=0\) and
\[
\frac{\lambda_j}{\alpha_j+\nu n_j}
-\frac{\lambda_j}{\alpha_j+\nu s_j^*}
=-\frac{\nu}{c_B^2}\delta_j
+\frac{\nu^2\delta_j^2}
{c_B^2(c_B\sqrt{\lambda_j}+\nu\delta_j)}.
\]
The linear terms cancel.  The condition also gives
\(c_B\sqrt{\lambda_j}+\nu\delta_j
\ge c_B\sqrt{\lambda_j}/2\), so summing the remainders gives
\[
0\le
\sum_{j=1}^r\frac{\lambda_j}{\alpha_j+\nu n_j}-\Psi_{\rm frac}(B)
\le
\frac{2\nu^2L^3H}{(\nu B+\alpha_\Sigma)^3}.
\]
Because the eigenvector probes are allowed in the arbitrary-direction problem
and the rounded allocation is feasible for the spectral problem,
\[
\Psi_{\rm frac}(B) \leq \Psi_{\rm all}(B) \leq \Psi_{\rm spec}(B) \leq
\sum_{j=1}^r \frac{\lambda_j}{\alpha_j+\nu n_j}.
\]
Combining this chain with the remainder estimate proves the absolute
bound. Since the large-budget condition gives
\[
\Psi_{\rm all}(B)\geq\Psi_{\rm frac}(B)
=\frac{L^2}{\nu B+\alpha_\Sigma},
\]
division by this lower bound proves the relative estimate.
\end{proof}

Thus the advantage of arbitrary directions vanishes with \(B\).  The eigenvector
restriction trades this vanishing gap for fewer settings and no feedback, which
can be far more costly than repeated shots, as shown in Subsection
\ref{subsec:common-access-cost}.

\section{Acquisition and Offline Inference}
\label{sec:acquisition-inference}

Algorithm \ref{alg:spectral-fidelity-design} fixes the probe states and their shot counts before any data are observed.  We now convert the acquired counts into an estimate of the unknown state and requested output, and then compare the resources needed to acquire the counts with those of adaptive competitors.

\subsection{Local estimator}
\label{subsec:local-estimator}

The exact likelihood in \eqref{eq:fidelity-likelihood} describes the measured
counts.  We convert them into an estimate in three steps: invert each probe
frequency into a raw local displacement, regularize the noisy displacements
with the prior, and reconstruct the state and requested output.  To distinguish
the parameters involved, write the unknown truth as
\[
\theta_\star=\theta_0+\sum_{j=1}^d b_{\star,j}v_j.
\]
Thus \(\theta_0\) is the known design centre, \(\theta_\star\) is the unknown
parameter, and \(b_{\star,j}\) is the true signed displacement from the centre
along \(v_j\); it is not a displacement from the probe.

Once \(G\) and \(Q\) are formed, constructing the fixed design requires one
dense \(d\)-dimensional generalized eigenproblem, costing \(O(d^3)\), while the
exact heap allocation costs \(O(d+B\log d)\) time and \(O(d)\) memory.

Consider an active spectral probe in direction \(v_j\), with success count
\(X_j\) from \(N_j>0\) direct-readout shots and observed frequency
\(\widehat p_j=X_j/N_j\).  The probe and the one-dimensional projective
geodesic through the centre are
\[
\phi_j=\cos\rho\,\psi_0+\sin\rho\,u_j,
\qquad
\psi_j(b)=\cos b\,\psi_0+\sin b\,u_j,
\]
where \(u_j\perp\psi_0\) and both vectors have unit norm.  Their overlap is
\[
\ip{\phi_j}{\psi_j(b)}
=\cos\rho\cos b+\sin\rho\sin b
=\cos(\rho-b),
\]
so the direct-readout success probability is
\begin{equation}
p_j(b)=\cos^2(\rho-b).
\label{eq:one-dimensional-probe-probability}
\end{equation}
On the branch \(0<\rho-b<\pi/2\), the cosine is positive and
\eqref{eq:one-dimensional-probe-probability} can be inverted uniquely:
\[
b=g_\rho(p):=\rho-\arccos\sqrt p.
\]
Replacing the unknown probability by its observed frequency gives the raw
data-only estimate
\begin{equation}
\widehat z_j
=g_\rho(\widehat p_j)
=\rho-\arccos\sqrt{X_j/N_j}.
\label{eq:raw-geodesic-inverse}
\end{equation}
For a spectral probe, the scalar being measured is \(z_j=b_{\star,j}\) in the
ideal one-dimensional model; the notation \(\widehat z_j\) emphasizes that
\eqref{eq:raw-geodesic-inverse} is the noisy inverse before prior
regularization.  At \(\rho=\pi/4\), frequency \(1/2\) gives
\(\widehat z_j=0\), a larger frequency indicates motion toward the probe, and
a smaller frequency indicates motion away from it.  The identity is exact on
the displayed geodesic.  For a curved multivariate state family, the tangent
agrees at the design centre, but applying the inverse separately to each probe
neglects cross-coordinate and higher-order terms and is therefore local.

The next result quantifies the finite-shot noise of this raw inverse.

\begin{proposition}[Finite-shot scale of the geodesic inverse]
\label{prop:geodesic-inverse-scale}
For one direct-readout spectral probe, suppose the local state lies on the
corresponding geodesic and \(p(b)=\cos^2(\rho-b)\), with
\(0<\rho-b<\pi/2\).  The estimator
\(\widehat z=\rho-\arccos\sqrt{X/N}\) satisfies
\[
\sqrt N\,(\widehat z-b)
\ \xrightarrow{\mathcal D}\
N(0,1/4).
\]
At the design centre \(b=0\), choosing \(\rho=\pi/4\) also makes its
second-order delta-method bias term vanish.
\end{proposition}

\begin{proof}
For \(g_\rho(p)=\rho-\arccos\sqrt p\),
\[
g_\rho'(p)=\frac{1}{2\sqrt{p(1-p)}},
\qquad
g_\rho'(p(b))=\frac1{\sin(2(\rho-b))}.
\]
The binomial central limit theorem gives
\(\sqrt N(X/N-p)\xrightarrow{\mathcal D}N(0,p(1-p))\).  Applying the delta
method to \(g_\rho\) gives asymptotic variance
\[
\{g_\rho'(p(b))\}^2p(b)\{1-p(b)\}
=\frac14.
\]
Moreover,
\[
g_\rho''(p)=\frac{2p-1}{4\{p(1-p)\}^{3/2}},
\]
so the second-order delta-method bias term vanishes at
\(p=1/2\), which is the design-centre probability when \(\rho=\pi/4\).
\end{proof}

Equivalently, the standard error is asymptotically \(1/(2\sqrt N)\).  The
variance \(1/(4N)\) is the inverse of the information in Proposition
\ref{prop:fidelity-information}.  Hence, near the design centre, the independent
spectral-probe estimates have the local Gaussian approximation
\[
\widehat z_j\mid b_{\star,j}\ \dot\sim
N\!\left(b_{\star,j},\frac{1}{4N_j}\right).
\]
The binomial likelihood \eqref{eq:fidelity-likelihood} remains finite-shot
exact.  Only the Gaussian approximation and the coordinatewise interpretation
of \(\widehat z_j\) are local or asymptotic.

For the closed-form update, approximate the centred local prior by
\[
b_j\sim N(0,\alpha_j^{-1}),
\qquad
\Lambda_0=\operatorname{diag}(\alpha_1,\ldots,\alpha_d).
\]
The observation precision is \(4N_j\), while the prior precision is
\(\alpha_j\).  Equivalently, the coordinatewise negative log posterior is,
up to an additive constant,
\[
\frac{\alpha_j}{2}b_j^2+2N_j(\widehat z_j-b_j)^2.
\]
Completing the square gives diagonal posterior precision \(P\) and unprojected
posterior mean (equivalently, mode) \(\bar b\):
\begin{equation}
P=\Lambda_0+4\sum_jN_j e_je_j^T,
\qquad
\bar b_j=\frac{4N_j}{\alpha_j+4N_j}\widehat z_j
\quad(N_j>0),
\qquad
\bar b_j=0\quad(N_j=0).
\label{eq:local-estimator}
\end{equation}
The multiplier \(4N_j/(\alpha_j+4N_j)\) is a precision weight in \([0,1)\).
It shrinks a noisy coordinate toward the prior centre, is small when prior
precision dominates, and approaches one as the measurement precision grows.
When \(N_j=0\), the data provide no update and the estimate remains at the prior
mean.  Because \(P\) is diagonal, the \(K_+\) inverse calibrations and
coordinatewise shrinkage cost \(O(K_++d)\) and require no dense inverse,
simulation, sampling, or optimization.  In the
uniform-ball experiments, \(\Lambda_0\) is the covariance-matched Gaussian
surrogate used by the local design, and the hard support is imposed by the
projection below.

Let
\[
\mathcal C_R:=\{b\in\R^d:\norm{b}_2\le R\}
\]
be the localization ball from Definition \ref{def:localization-radius}, and
write the task metric in the generalized coordinates as
\[
\widetilde Q=\operatorname{diag}(\lambda_1,\ldots,\lambda_d).
\]
We enforce physical localization using the metric of the requested task.  Define
the selected task-metric projection by
\begin{equation}
\Pi_{R,Q}(\bar b)
\in
\argmin_{b\in\mathcal C_R}
(b-\bar b)^T\widetilde Q(b-\bar b),
\label{eq:task-localization-projection}
\end{equation}
choosing among multiple minimizers the one with smallest Euclidean norm.  This
tie-break is needed because a low-rank task has \(\ker(\widetilde Q)\ne\{0\}\).
It sets task-null coordinates to zero and makes the selected estimate unique.

The projection remains inexpensive.  Let \(\bar b_+\) equal \(\bar b\) on
coordinates with \(\lambda_j>0\) and zero elsewhere.  If
\(\norm{\bar b_+}_2\le R\), then
\(\Pi_{R,Q}(\bar b)=\bar b_+\).  Otherwise,
\begin{equation}
\{\Pi_{R,Q}(\bar b)\}_j
=
\begin{cases}
\displaystyle\frac{\lambda_j}{\lambda_j+\mu}\bar b_j,
& \lambda_j>0,\\[2mm]
0, & \lambda_j=0,
\end{cases}
\label{eq:task-projection-coordinates}
\end{equation}
where the unique \(\mu>0\) solves
\begin{equation}
\sum_{j:\lambda_j>0}
\left(\frac{\lambda_j}{\lambda_j+\mu}\bar b_j\right)^2
=R^2.
\label{eq:task-projection-multiplier}
\end{equation}
The left-hand side is continuous and strictly decreasing on the relevant
coordinates, so \(\mu\) is obtained by a scalar root search.  Let \(L_\mu=0\)
when projection is inactive and otherwise let \(L_\mu\) denote the number of
root-search iterations.  Each iteration costs \(O(d)\), so projection costs
\(O((L_\mu+1)d)\); bracketed bisection gives
\(L_\mu=O(\log(1/\varepsilon_\mu))\) for multiplier tolerance
\(\varepsilon_\mu\).  No multidimensional optimization is needed.  When
\(Q=G\), all \(\lambda_j=1\), and
\eqref{eq:task-projection-coordinates} reduces to ordinary radial projection
onto \(\mathcal C_R\).

The following guarantee is a direct finite-dimensional, single-ball instance
of the convex projection principle used in
\cite{hay_sharon_anchor_projection}.

\begin{proposition}[Safe task-metric localization projection]
\label{prop:safe-localization-projection}
Let \(\widetilde Q\succeq0\).  Suppose the pre-data localization assumption is
valid, so \(b_\star\in\mathcal C_R\), and set
\(\widehat b=\Pi_{R,Q}(\bar b)\).  Then
\begin{equation}
\norm{\widehat b-b_\star}_{\widetilde Q}^2
\le
\norm{\bar b-b_\star}_{\widetilde Q}^2
-\norm{\bar b-\widehat b}_{\widetilde Q}^2
\le
\norm{\bar b-b_\star}_{\widetilde Q}^2,
\label{eq:localization-projection-guarantee}
\end{equation}
where \(\norm{x}_{\widetilde Q}^2=x^T\widetilde Qx\).  Thus projection cannot
increase the local quadratic task loss and strictly decreases it whenever
\(\norm{\bar b-\widehat b}_{\widetilde Q}>0\).
\end{proposition}

\begin{proof}
Let
\(E_+=\operatorname{span}\{e_j:\lambda_j>0\}\), on which
\(\ip{x}{y}_{\widetilde Q}=x^T\widetilde Qy\) is an inner product, and let
\(P_+\) be the Euclidean projection onto \(E_+\).  Since
\(b_\star\in\mathcal C_R\), its task-relevant part satisfies
\(P_+b_\star\in\mathcal C_R\cap E_+\).  The minimum-norm convention in
\eqref{eq:task-localization-projection} gives \(\widehat b\in E_+\), and
\(\widehat b\) is precisely the \(\widetilde Q\)-metric projection of
\(P_+\bar b\) onto the closed convex set \(\mathcal C_R\cap E_+\).

Now apply the Hilbert-space projection argument in Theorem 3.6 of
\cite{hay_sharon_anchor_projection}.  Its proof uses only closed convexity, and
applies here with the identifications
\[
(\mathcal F,\ip{\cdot}{\cdot}_\Xi)
=(E_+,\ip{\cdot}{\cdot}_{\widetilde Q}),
\quad
S=\mathcal C_R\cap E_+,
\quad
f=P_+b_\star,
\quad
g=P_+\bar b,
\quad
h=\widehat b.
\]
The resulting Pythagorean projection inequality is exactly
\eqref{eq:localization-projection-guarantee}, because task-null components do
not contribute to the \(\widetilde Q\)-seminorm.  If \(E_+=\{0\}\), both sides
are zero and the claim is immediate.
\end{proof}

Unlike the earlier anchor certificates, Proposition
\ref{prop:safe-localization-projection} is conditional on the pre-data
localization assumption \(b_\star\in\mathcal C_R\); the measurement outcomes do not certify that assumption. However, under this condition, the result is valid for every positive-semidefinite task metric, including low-rank tasks and
 full-state specialization \(Q=G\).

Finally, transform the feasible spectral coordinates back to the model
parameter and evaluate the requested output:
\[
\widehat\theta=\theta_0+\sum_j\widehat b_jv_j,
\qquad
\widehat A=A(\widehat\theta).
\]
Thus all post-acquisition coordinate algebra costs
\(O(K_++(L_\mu+1)d)\), followed by one model and task evaluation.  This is a
plug-in estimate: the implementation evaluates \(A\) once at the estimated
parameter.  For nonlinear \(A\), it generally differs from the posterior
average \(\mathbb E[A(\theta)\mid X]\).  Even before localization projection,
evaluating \(A\) at the Gaussian posterior mean generally does not commute with
posterior averaging.

\subsection{Common access model and acquisition cost}
\label{subsec:common-access-cost}

Having described how the fixed-batch counts are converted into an estimate, we
now compare the resources required to acquire those counts.
The proposed method, SGQT, OSGQT, and PAQT all query fidelity to prepared
proposal states.  We give every method the same direct projective readout:
applying \(U_{\phi_i}^{\dagger}\) to a target copy and accepting the all-zero
outcome succeeds with probability \(\abs{\ip{\phi_i}{\psi}}^2\).  Let \(T\) be
the number of paired adaptive iterations, \(B\) the number of target copies,
and \(K_+\) the number of active fixed probes.  Let \(s\) be the cost of one
target copy and shot, \(a\) the cost of creating, compiling, submitting, and
calibrating a distinct setting, and \(r\) the cost of one feedback round.  The
model compares these common access costs without assuming equal circuit depth.
If preliminary measurements are used to obtain the pilot, their copy, setting,
and feedback costs must be added to the acquisition budget.

\begin{table}[H]
\centering
\scriptsize
\caption{Acquisition costs under the common projective-fidelity access model.
In all reported comparisons, \(K_+\ll T\).  In the timing scenarios below,
\(s\ll a\ll r\).  The costs
exclude pre-batch design and method-specific classical inference.}
\label{tab:access-cost}
\begin{tabular}{@{}lcccc@{}}
\toprule
Method & Feedback rounds & Settings & Copies & Acquisition cost\\
\midrule
Spectral fixed batch & \(0\) & \(K_+\) & \(B\) & \(Bs+K_+a\)\\
SGQT & \(T\) & \(2T\) & \(B\) & \(Bs+2Ta+Tr\)\\
OSGQT & \(T\) & \(2T\) & \(B\) & \(Bs+2Ta+Tr\)\\
PAQT & \(T\) & \(2T\) & \(B\) & \(Bs+2Ta+Tr\)\\
\bottomrule
\end{tabular}
\end{table}

Every competitor is tested at four \(T\)'s and granted its post hoc best value.
These 25--150-shot schedules have \(K_+\le2,T\ge64\) for the task and
\(K_+\le15,T\ge512\) for full state, hence \(K_+\ll T\) in both.  At equal
\(B\), adaptive-minus-fixed acquisition costs
\((2T-K_+)a+Tr>0\) whenever \(a>0\) or \(r>0\).

A latency-aware VQA study uses illustrative times of \(10\,\mu\mathrm{s}\) per
shot, \(100\,\mathrm{ms}\) per circuit switch, and \(4\,\mathrm{s}\) per
communication round \cite{ito_latency_aware_2023}.  Identifying these with
\(s,a,r\) and normalizing by the shot time gives
\(s:a:r=1:10000:400000\).  We use this as an illustrative acquisition
scenario, not as a complete end-to-end timing claim.

\begin{remark}[Classical operation counts]
\label{rem:classical-cost}
The costs of the Greedy spectral design and estimator are stated beside their
corresponding steps in Subsection \ref{subsec:local-estimator}.  Forming \(G\)
and \(Q\) from the pilot derivatives is model dependent and is performed once.
After they are formed, the one-off design stage solves a dense generalized
eigenproblem in \(O(d^3)\) time and performs the exact allocation in
\(O(d+B\log d)\) time and \(O(d)\) memory.  For a fixed pilot, task, prior, and
budget, this design can be reused across repeated acquisitions.  Each acquired
batch then requires \(O(K_++(L_\mu+1)d)\) coordinate work, followed by one model
and task evaluation.  Thus, after metric formation, the total cost for the
first batch is
\(O(d^3+B\log d+K_++(L_\mu+1)d)\), while later batches omit the first two
terms.

PAQT instead evaluates \(2T\) count-block likelihoods over \(N_{\rm p}\)
particles for every batch, costing \(O(N_{\rm p}T)\) before posterior averaging,
with additional particle-state evaluations whenever Liu--West resampling is
triggered.
In the reported schedules, \(N_{\rm p}=8000\) and \(T\ge64\), so PAQT requires
at least \(1.024\times10^6\) particle--setting likelihood evaluations.  Hence,
at the reported dimensions and schedules, the proposed method has substantially
lower counted classical work even when its one-off eigensolve and allocation
are included; reuse further lowers its amortized cost.  This comparison does
not equate different primitive operations or assign a universal cost to metric
formation and final model evaluation.
\end{remark}

Cheaper direct-task or shadow measurements may be preferable under other
primitives \cite{aaronson_shadow_tomography,
huang_kueng_preskill_classical_shadows}; this comparison is restricted to
fidelity-to-proposal access and equal \(B\).  ABQT is also omitted because it
searches after every measurement for full-state information gain
\cite{huszar_houlsby_adaptive_bayesian}.  PAQT is the closest practical
comparator: it retains Bayesian inference with the cheaper SGQT/SPSA selection
rule \cite{granade_paqt_2016}.

\section{Numerical Experiments}
\label{sec:experiments}

\subsection{Schmidt-spectrum task}

We study the ground state of a six-qubit disordered transverse-field Ising
family
\begin{equation}
H(\theta)=H_0+\sum_{j=1}^{d}\theta_jV_j.
\label{eq:ising-family}
\end{equation}
The nested families \(d=6,12,17\) successively add local \(X\) fields, local
\(Z\) fields, and nearest-neighbor \(ZZ\) couplings.  A relative
Fubini--Study metric cutoff of \(10^{-8}\) gives local dimensions
\(d_{\rm loc}=6,12,15\).  All three families share the same nominal ground
state, isolating the effect of enlarging the perturbation space.  Appendix
\ref{app:benchmark-contracts} records the exact model and key stability checks.

The primary task is the ordered Schmidt spectrum across the odd--even \(3|3\)
cut \(A=\{0,2,4\}\), \(A^c=\{1,3,5\}\).  If
\[
\rho_A=\Tr_{\{1,3,5\}}\lvert\psi\rangle\langle\psi\rvert,
\qquad
p_1\ge\cdots\ge p_8,
\]
then
\[
A(\theta)=\tau(\psi(\theta))=(p_1,\ldots,p_8)^T.
\]
This nonlinear output is not the expectation of a fixed single-copy
observable.  We differentiate it only at the nondegenerate pilot; numerical
endpoints use sorted spectra.

At the nominal state, centered finite differences give the horizontal state
derivatives and task Jacobian \(J\).  Because the reported score is the raw
Euclidean spectrum loss, its local quadratic pullback is
\[
Q=J^TJ.
\]
Thus the theorem-aligned local loss and the linearization of the reported score
use the same norm.  The numerical task ranks are \(5,6,7\) for
\(d=6,12,17\), respectively, and the strongly anisotropic eigenvalues in
Table \ref{tab:task-spectrum} motivate nonuniform allocation.

\begin{table}[H]
\centering
\scriptsize
\setlength{\tabcolsep}{3pt}
\caption{Positive generalized task eigenvalues of the raw Euclidean local
Schmidt-spectrum map.}
\label{tab:task-spectrum}
\resizebox{\textwidth}{!}{%
\begin{tabular}{ccrrrrrrr}
\toprule
\(d\) & rank & \(\lambda_1\) & \(\lambda_2\) & \(\lambda_3\) &
\(\lambda_4\) & \(\lambda_5\) & \(\lambda_6\) & \(\lambda_7\)\\
\midrule
6 & 5 & \(3.5315{\times}10^{-1}\) & \(8.9105{\times}10^{-3}\) &
\(2.4992{\times}10^{-4}\) & \(1.1920{\times}10^{-6}\) &
\(5.9516{\times}10^{-10}\) & -- & --\\
12 & 6 & \(5.9710{\times}10^{-1}\) & \(2.5814{\times}10^{-2}\) &
\(3.0562{\times}10^{-4}\) & \(1.4820{\times}10^{-4}\) &
\(4.5226{\times}10^{-7}\) & \(2.0249{\times}10^{-8}\) & --\\
17 & 7 & \(7.5166{\times}10^{-1}\) & \(1.3680{\times}10^{-1}\) &
\(4.7563{\times}10^{-3}\) & \(3.2286{\times}10^{-4}\) &
\(1.7198{\times}10^{-5}\) & \(2.4900{\times}10^{-7}\) &
\(3.5132{\times}10^{-8}\)\\
\bottomrule
\end{tabular}
}
\end{table}

Following Definition \ref{def:localization-radius}, we choose \(R=0.02\)
before generating the truths, rather than estimating it from the outcomes.
Truths and structured priors are uniform on \(\norm{b}_2\le R\) in the
generalized coordinates of \eqref{eq:task-coordinate-loss}.  Covariance
matching gives
\[
\mathbb E[bb^T]=\frac{R^2}{d_{\rm loc}+2}I,
\qquad
\alpha=\frac{d_{\rm loc}+2}{R^2}
\]
for the isotropic Gaussian precision.  We use the prefix S- (``structured'')
for an adaptive competitor that operates in the local Hamiltonian coordinates
and is constrained to the localization ball, rather than searching over
arbitrary Hilbert-space states.  Our S-PAQT adaptation starts at the common
pilot, uses uniform-ball particles and Liu--West resampling, and otherwise
retains the published gains \cite{granade_paqt_2016}; Appendix
\ref{app:benchmark-contracts} records the support and probe conventions.  For
S-SGQT and S-OSGQT, the probes and
updates are parameterized in the \(d_{\rm loc}\) Hamiltonian coordinates, and
each updated iterate is projected onto the ball.  The S-variants therefore
receive the same pre-data structural information available to our method,
including the local family, retained dimension \(d_{\rm loc}\), and radius
\(R\).  This strengthens the adaptive baselines and makes the comparison fairer:
an observed advantage cannot be attributed merely to withholding the known
model and localization support from the competitors.  Locally, the boundary
corresponds to
\(1-F\simeq R^2=4\times10^{-4}\).  For scale, this is near the numerical
midpoint between the single-qubit state-preparation error estimated by Harty
et al.\ \((\approx2\times10^{-4})\) and that measured by Leu et al.\
\((6.1(11)\times10^{-4})\)
\cite{harty_high_fidelity_2014,leu_polarisation_insensitive_2024}.  These
single-qubit figures do not determine our six-qubit radius; they only place its
boundary between two published experimental scales.  The supplementary
materials independently vary truth displacement and the assumed radius from
\(0.01\) to \(0.16\), distinguishing failure caused by leaving the pilot from
failure caused by an overly broad assumed radius.

Every fixed probe uses direct projective readout at \(\rho=\pi/4\), so
\(\nu_j=4\).  We use
\[
B\in\{19200,38400,76800,153600\}.
\]
Table \ref{tab:task-allocations} gives the integer allocations, and Appendix \ref{app:benchmark-contracts} gives the exact benchmark specification.
\begin{table}[htp]
\centering
\caption{Greedy spectral task allocations.  \(K_+\) is the number of nonzero fixed
settings.}
\label{tab:task-allocations}
\begin{tabular}{r@{\hspace{14pt}}rrrr}
\toprule
\(d\) & \(B\) & \(n_1\) & \(n_2\) & \(K_+\)\\
\midrule
6  & 19200  & 19200  & 0     & 1\\
6  & 38400  & 36766  & 1634  & 2\\
6  & 76800  & 69902  & 6898  & 2\\
6  & 153600 & 136175 & 17425 & 2\\
\midrule
12 & 19200  & 19200  & 0     & 1\\
12 & 38400  & 37528  & 872   & 2\\
12 & 76800  & 69318  & 7482  & 2\\
12 & 153600 & 132898 & 20702 & 2\\
\midrule
17 & 19200  & 17729  & 1471  & 2\\
17 & 38400  & 31187  & 7213  & 2\\
17 & 76800  & 58104  & 18696 & 2\\
17 & 153600 & 111938 & 41662 & 2\\
\bottomrule
\end{tabular}
\end{table}

Only one or two retained directions are active, so the all-active rate in Theorem~\ref{thm:spectral_cost} is not invoked in these experiments. By~\eqref{eq:fractional_certificate}, the relative improvement available from arbitrary directions is at most $\Psi_{\rm spec}(B)/\Psi_{\rm frac}(B)-1$, whose maximum over the twelve task cells is less than $3\times10^{-10}$. Thus, the sparse allocations are the exact marginal-gain output, not a post hoc truncation; at the largest budget, the remaining unmeasured positive directions contribute at most $5.2\%$ of the local surrogate.

We compare our Greedy spectral design with S-PAQT, S-SGQT, and S-OSGQT under
the same family, pilot, radius, copy budget, and projective-fidelity access.  The adaptive
methods use fidelity rather than task gradients.  Because they do not specify
a unique round count \(T\), we test the same four budget-scaled values for each
method and report the smallest observed mean.  This post hoc choice is
deliberately favorable to the adaptive baselines.  Appendix
\ref{app:benchmark-contracts} provides adaptive schedules, endpoints, and finite-particle controls.

The comparison therefore concerns complete acquisition--inference pipelines; the ablation below isolates direction and allocation effects.  Means pair the same 30 truths, and error bars are Monte Carlo standard errors for
\[
L_{\rm task}=\norm{\widehat\tau-\tau}_2^2.
\]
Across the \(3\times4\times4\times3=144\)
dimension--budget--round--method cells, Greedy spectral has the lower mean in all 144.
After selecting each adaptive method's best \(T\), Greedy spectral remains lower in all
12 dimension--budget cells.  At \(B=153600\), it reduces the best adaptive
mean by \(71.6\%\), \(87.4\%\), and \(87.4\%\) for \(d=6,12,17\),
respectively.  Of the 36 method-specific oracle comparisons, 34 paired
intervals exclude zero; the S-PAQT intervals at \(d=12\) and
\(B\in\{19200,38400\}\) are unresolved.

\begin{table}[t]
\centering
\scriptsize
\caption{Task MSE on 30 paired truths.  In every dimension--budget cell, each
adaptive method is independently granted its smallest observed mean over four
budget-scaled round counts \(T\), selected post hoc using the same evaluation
truths, with the selected \(T\) shown in parentheses.  This method- and
cell-specific oracle tuning is deliberately competitor-favorable and would not
be available in prospective deployment.  Greedy spectral requires no choice of
\(T\).}
\label{tab:task-round-sensitivity}
\begin{tabular}{rrcccc}
\toprule
\(d\) & \(B\) & Greedy spectral & \(\min_T\) S-PAQT & \(\min_T\) S-SGQT & \(\min_T\) S-OSGQT\\
\midrule
6 & 19200
& \(\mathbf{6.01{\times}10^{-6}}\) & \(1.24{\times}10^{-5}\;(64)\)
& \(4.00{\times}10^{-5}\;(384)\) & \(3.66{\times}10^{-5}\;(192)\)\\
6 & 38400
& \(\mathbf{2.87{\times}10^{-6}}\) & \(9.47{\times}10^{-6}\;(768)\)
& \(3.36{\times}10^{-5}\;(192)\) & \(3.28{\times}10^{-5}\;(192)\)\\
6 & 76800
& \(\mathbf{1.29{\times}10^{-6}}\) & \(4.79{\times}10^{-6}\;(256)\)
& \(3.45{\times}10^{-5}\;(384)\) & \(3.53{\times}10^{-5}\;(384)\)\\
6 & 153600
& \(\mathbf{8.27{\times}10^{-7}}\) & \(2.91{\times}10^{-6}\;(768)\)
& \(3.36{\times}10^{-5}\;(1536)\) & \(3.90{\times}10^{-5}\;(768)\)\\
\addlinespace
12 & 19200
& \(\mathbf{6.77{\times}10^{-6}}\) & \(1.11{\times}10^{-5}\;(384)\)
& \(2.98{\times}10^{-5}\;(96)\) & \(2.80{\times}10^{-5}\;(64)\)\\
12 & 38400
& \(\mathbf{3.84{\times}10^{-6}}\) & \(9.07{\times}10^{-6}\;(384)\)
& \(3.22{\times}10^{-5}\;(384)\) & \(2.90{\times}10^{-5}\;(192)\)\\
12 & 76800
& \(\mathbf{2.55{\times}10^{-6}}\) & \(7.46{\times}10^{-6}\;(768)\)
& \(3.70{\times}10^{-5}\;(256)\) & \(2.08{\times}10^{-5}\;(256)\)\\
12 & 153600
& \(\mathbf{9.48{\times}10^{-7}}\) & \(7.55{\times}10^{-6}\;(512)\)
& \(2.63{\times}10^{-5}\;(3072)\) & \(2.96{\times}10^{-5}\;(768)\)\\
\addlinespace
17 & 19200
& \(\mathbf{1.06{\times}10^{-5}}\) & \(2.42{\times}10^{-5}\;(64)\)
& \(6.14{\times}10^{-5}\;(384)\) & \(3.68{\times}10^{-5}\;(64)\)\\
17 & 38400
& \(\mathbf{6.19{\times}10^{-6}}\) & \(2.23{\times}10^{-5}\;(384)\)
& \(6.18{\times}10^{-5}\;(768)\) & \(3.91{\times}10^{-5}\;(384)\)\\
17 & 76800
& \(\mathbf{3.20{\times}10^{-6}}\) & \(2.04{\times}10^{-5}\;(384)\)
& \(6.49{\times}10^{-5}\;(1536)\) & \(3.60{\times}10^{-5}\;(384)\)\\
17 & 153600
& \(\mathbf{2.23{\times}10^{-6}}\) & \(1.77{\times}10^{-5}\;(512)\)
& \(4.81{\times}10^{-5}\;(3072)\) & \(3.47{\times}10^{-5}\;(768)\)\\
\bottomrule
\end{tabular}
\end{table}
\begin{figure}[t]
\centering
\includegraphics[width=0.92\textwidth]{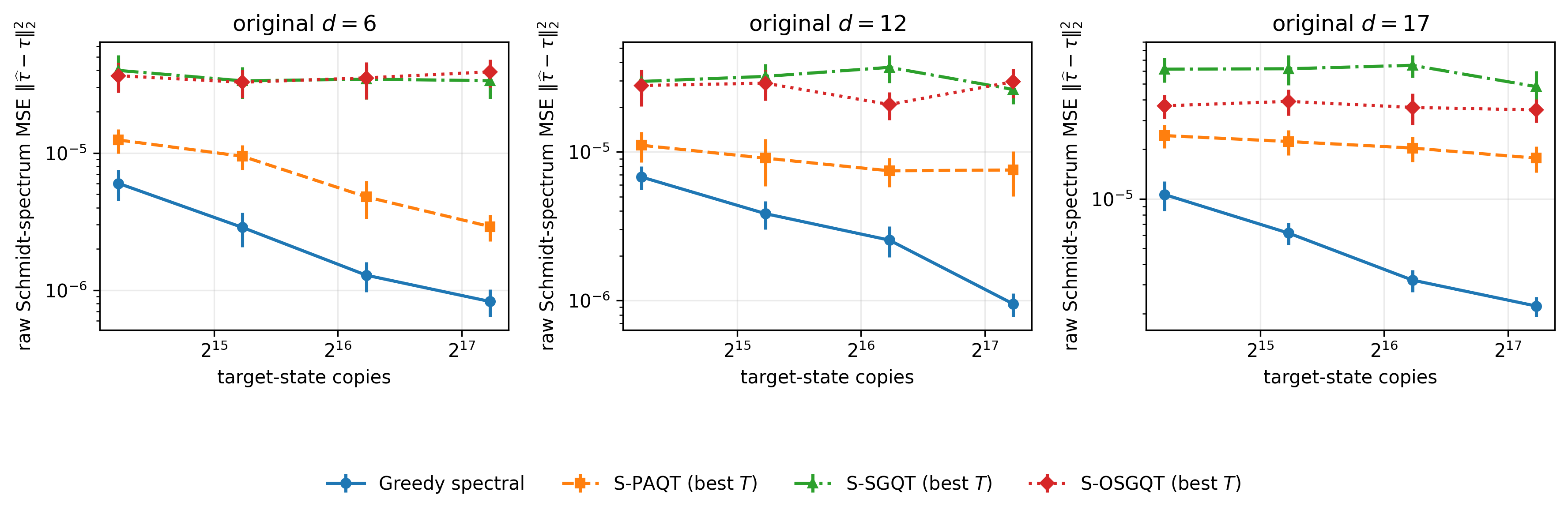}
\caption{Mean raw Schmidt-spectrum MSE versus target-copy budget.  Adaptive
points use the best observed \(T\) recorded in Table
\ref{tab:task-round-sensitivity}; Greedy spectral is independent of \(T\).  Vertical scales are logarithmic, and error bars are Monte Carlo standard errors.}
\label{fig:task-benchmark}
\end{figure}

\begin{table}[H]
\centering
\scriptsize
\setlength{\tabcolsep}{4.2pt}
\caption{Task acquisition summary over the four copy budgets.  Adaptive ranges
include the method-specific selected schedules in Table
\ref{tab:task-round-sensitivity}.  Cost ratios use
\(s:a:r=1:10000:400000\) in Table \ref{tab:access-cost}; each adaptive round
uses two settings, and ratio endpoints are rounded upward to integers.}
\label{tab:task-acquisition-summary}
\begin{tabular}{@{}cccc@{}}
\toprule
\(d\) & Fixed \(K_+\) & Adaptive rounds \(T\) &
\(C_{\rm adaptive}/C_{\rm fixed}\)\\
\midrule
6  & \(1\)--\(2\) & \(64\)--\(1536\) & \(922\)--\(5524\)\\
12 & \(1\)--\(2\) & \(64\)--\(3072\) & \(922\)--\(7434\)\\
17 & \(2\) & \(64\)--\(3072\) & \(687\)--\(7434\)\\
\bottomrule
\end{tabular}
\end{table}
\vspace{-\baselineskip}


Across all 12 tested cases, our method has lower mean error while using only
one or two fixed settings and no feedback.  Based on the latency estimates of
Ito et al.~\cite{ito_latency_aware_2023}, adaptive acquisition costs
\(687\)--\(7434\) times more; classical costs are discussed in
Remark~\ref{rem:classical-cost}.  To favor the adaptive methods, we chose \(T\)
and the reported method using the evaluation truths, so the intervals do not
account for this selection.  The S-PAQT results are also empirical estimates
based on a finite number of particles.  At \(B=153600\), S-PAQT ends with a
mean ESS of \(5882\)--\(6065\) out of 8000 particles and averages
\(8.7\)--\(9.8\) resampling events per trial.  The next ablation tests the
design with finite-binomial task MSE.

\subsection{Allocation and direction-selection ablations}

We first isolate allocation within the task spectral subspace.  Equal spectral
allocation uses the same \(r\) retained probes as our Greedy spectral design, while
the \(\lambda\)-proportional allocation assigns \(n_j\propto\lambda_j\) once
before acquisition and does not update marginal gains.  Three direction
controls complete the comparison:
coordinate and random assign shots equally to, respectively, the \(r\)
whitened coordinate axes with largest task Rayleigh quotients and \(r\)
fixed random orthonormal directions; nuisance uses as many task-null directions
as are available (one, six, and seven for \(d=6,12,17\)).  All six designs are fixed before acquisition, use
\(\rho=\pi/4\), and are evaluated on the same 300 truths at \(R=0.02\).

We report the objective proved in Theorem \ref{thm:greedy-optimality} for spectral-probe designs, together with the empirical local quadratic error
\((\widehat b-b)^TQ(\widehat b-b)\) and raw physical task MSE for every design.

\begin{table}[H]
\centering
\scriptsize
\caption{Allocation and direction-selection endpoints at \(B=153600\) for the endpoint dimensions \(d=6\) and \(d=17\).  The deterministic \(\Psi_r\) is reported only for spectral-probe designs; empirical columns are means over the same 300 truths. Our proposed method is labeled Greedy spectral (ours). Bold marks the smallest reported value within each dimension and metric.}
\label{tab:compact-ablation}
\begin{tabular}{clrrr}
\toprule
\(d\) & Design & Predicted \(\Psi_r\) & Empirical local error & Raw task MSE\\
\midrule
6 & \textbf{Greedy spectral (ours)} & \(\mathbf{7.3727{\times}10^{-7}}\)
  & \(\mathbf{1.0724{\times}10^{-6}}\) & \(\mathbf{8.71{\times}10^{-7}}\)\\
  & Equal & \(2.5358{\times}10^{-6}\) & \(3.0583{\times}10^{-6}\)
  & \(3.01{\times}10^{-6}\)\\
  & \(\lambda\)-proportional & \(8.3671{\times}10^{-7}\)
  & \(1.1076{\times}10^{-6}\) & \(9.42{\times}10^{-7}\)\\
  & Coordinate & --- & \(2.9072{\times}10^{-6}\) & \(2.88{\times}10^{-6}\)\\
  & Random & --- & \(3.3453{\times}10^{-6}\) & \(3.25{\times}10^{-6}\)\\
  & Nuisance & --- & \(1.9113{\times}10^{-5}\) & \(1.90{\times}10^{-5}\)\\
\midrule
17 & \textbf{Greedy spectral (ours)} & \(\mathbf{2.3072{\times}10^{-6}}\)
   & \(\mathbf{2.6676{\times}10^{-6}}\) & \(\mathbf{2.24{\times}10^{-6}}\)\\
   & Equal & \(6.8592{\times}10^{-6}\) & \(6.8344{\times}10^{-6}\)
   & \(6.43{\times}10^{-6}\)\\
   & \(\lambda\)-proportional & \(2.4575{\times}10^{-6}\)
   & \(2.6833{\times}10^{-6}\) & \(2.33{\times}10^{-6}\)\\
   & Coordinate & --- & \(6.4362{\times}10^{-6}\) & \(6.18{\times}10^{-6}\)\\
   & Random & --- & \(1.3728{\times}10^{-5}\) & \(1.35{\times}10^{-5}\)\\
   & Nuisance & --- & \(1.9933{\times}10^{-5}\) & \(1.98{\times}10^{-5}\)\\
\bottomrule
\end{tabular}
\end{table}
\vspace{-\baselineskip}
Within the spectral family, our Greedy spectral design has the smallest predicted surrogate in
every dimension, exactly as the theorem requires.  The direction controls are
also worse on both empirical endpoints displayed.  Its raw-MSE reduction
relative to equal allocation is \(71.1\%\), \(73.6\%\), and \(65.1\%\) for
\(d=6,12,17\).  The \(\lambda\)-proportional rule is close but is worse in the
empirical endpoints displayed as well as in the proved surrogate.
The finite-binomial inverse, curved state and task maps, projection, and Monte
Carlo variation are absent from \(\Psi_r\); the empirical ordering is therefore
reported separately from the exact surrogate guarantee.

The direction-control rows in Table \ref{tab:compact-ablation} show that
coordinate and random probes
spend precision outside the leading task eigenspace, while nuisance probes
approach the prior task loss. Thus selecting the task spectral subspace is the
dominant effect; diminishing-return allocation adds an improvement within that
subspace.

\subsection{Full-state specialization}
\label{sec:full-state}

The same Greedy spectral method specializes to full-state tomography by setting \(Q=G\) on all \(d_{\rm loc}\) numerically retained directions.  
All whitened task eigenvalues equal one. Since the prior is isotropic and
$B=153600$ is divisible by both retained dimensions
$d_{\rm loc}\in\{6,15\}$, the balanced allocation across the retained probes
attains $\Psi_{\rm frac}(B)$. Hence
\[
\Psi_{\rm spec}(B)=\Psi_{\rm all}(B)=\Psi_{\rm frac}(B)
\]
for the local full-state surrogate. The score is the raw full-state infidelity
\[
L_{\rm state}
=
1-\langle\psi_{\rm true}|\widehat\rho|\psi_{\rm true}\rangle.
\]
At each \(R\in\{0.01,0.02,0.03,0.04,0.08,0.16\}\), all methods use the
same 30 unit-ball draws, scaled by \(R\).  Each adaptive method receives the common grid
\(T\in\{512,768,1536,3072\}\) and is summarized by its smallest observed
mean, a competitor-favorable post hoc choice.  S-PAQT reports the Bayesian
mean density operator from all counts.  Its 500-particle Liu--West filter
remains in the structured family and assumed radius.  Broad-radius rankings
are exploratory because of the small particle cloud and post hoc round
selection.

\begin{figure}[H]
\centering
\includegraphics[width=0.8\textwidth]{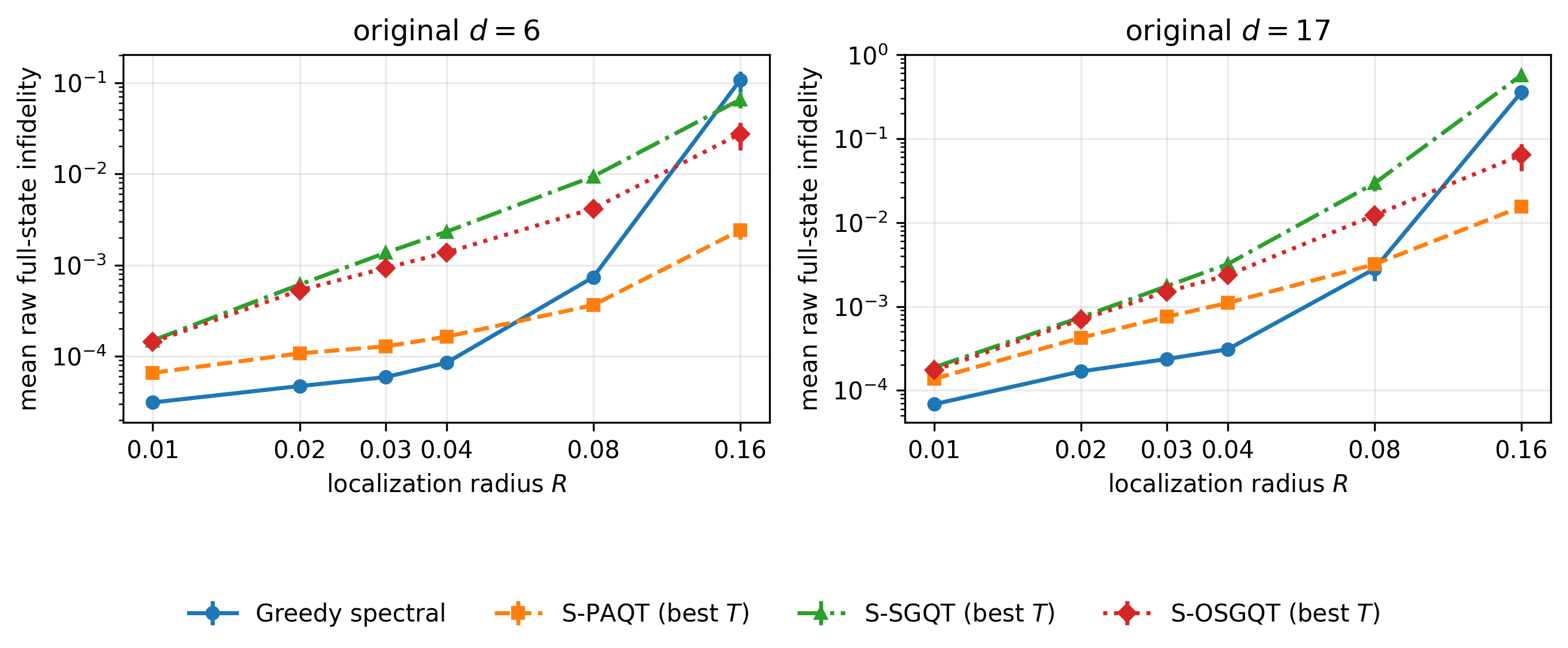}
\caption{Mean raw full-state infidelity versus localization radius.  Adaptive
curves use each method's best observed \(T\) from Table
\ref{tab:full-state-round-sensitivity}.  Both axes are logarithmic and error
bars are Monte Carlo standard errors.}
\label{fig:full-state-benchmark}
\end{figure}

Table \ref{tab:full-state-round-sensitivity} reports every method's selected
round count and mean, and Figure \ref{fig:full-state-benchmark} shows the same
radius dependence with Monte Carlo standard errors.

\begin{table}[H]
\centering
\scriptsize
\setlength{\tabcolsep}{3.5pt}
\caption{Raw full-state infidelity on 30 paired truths at \(B=153600\).  In
every dimension--radius cell, each adaptive method is independently granted its
smallest observed mean over four round counts \(T\), selected post hoc using the
same evaluation truths, with the selected \(T\) shown in parentheses.  This
method- and cell-specific oracle tuning is deliberately competitor-favorable
and would not be available in prospective deployment.  Greedy spectral requires
no choice of \(T\).  Bold marks the smallest mean in each dimension--radius
cell.}
\label{tab:full-state-round-sensitivity}
\begin{tabular}{rrcccc}
\toprule
\(d\) & \(R\) & Greedy spectral & \(\min_T\) S-PAQT & \(\min_T\) S-SGQT & \(\min_T\) S-OSGQT\\
\midrule
6 & 0.01 & \(\mathbf{3.11{\times}10^{-5}}\) & \(6.57{\times}10^{-5}\;(1536)\)
& \(1.49{\times}10^{-4}\;(3072)\) & \(1.43{\times}10^{-4}\;(3072)\)\\
6 & 0.02 & \(\mathbf{4.70{\times}10^{-5}}\) & \(1.07{\times}10^{-4}\;(3072)\)
& \(6.10{\times}10^{-4}\;(1536)\) & \(5.27{\times}10^{-4}\;(3072)\)\\
6 & 0.03 & \(\mathbf{5.89{\times}10^{-5}}\) & \(1.28{\times}10^{-4}\;(3072)\)
& \(1.36{\times}10^{-3}\;(1536)\) & \(9.27{\times}10^{-4}\;(512)\)\\
6 & 0.04 & \(\mathbf{8.47{\times}10^{-5}}\) & \(1.64{\times}10^{-4}\;(1536)\)
& \(2.32{\times}10^{-3}\;(1536)\) & \(1.37{\times}10^{-3}\;(512)\)\\
6 & 0.08 & \(7.35{\times}10^{-4}\) & \(\mathbf{3.63{\times}10^{-4}}\;(1536)\)
& \(9.34{\times}10^{-3}\;(1536)\) & \(4.15{\times}10^{-3}\;(512)\)\\
6 & 0.16 & \(1.06{\times}10^{-1}\) & \(\mathbf{2.40{\times}10^{-3}}\;(512)\)
& \(6.50{\times}10^{-2}\;(512)\) & \(2.74{\times}10^{-2}\;(512)\)\\
\midrule
17 & 0.01 & \(\mathbf{6.85{\times}10^{-5}}\) & \(1.37{\times}10^{-4}\;(512)\)
& \(1.88{\times}10^{-4}\;(3072)\) & \(1.74{\times}10^{-4}\;(3072)\)\\
17 & 0.02 & \(\mathbf{1.68{\times}10^{-4}}\) & \(4.24{\times}10^{-4}\;(512)\)
& \(7.44{\times}10^{-4}\;(768)\) & \(6.96{\times}10^{-4}\;(3072)\)\\
17 & 0.03 & \(\mathbf{2.36{\times}10^{-4}}\) & \(7.53{\times}10^{-4}\;(1536)\)
& \(1.75{\times}10^{-3}\;(768)\) & \(1.50{\times}10^{-3}\;(512)\)\\
17 & 0.04 & \(\mathbf{3.09{\times}10^{-4}}\) & \(1.10{\times}10^{-3}\;(1536)\)
& \(3.20{\times}10^{-3}\;(3072)\) & \(2.36{\times}10^{-3}\;(512)\)\\
17 & 0.08 & \(\mathbf{2.81{\times}10^{-3}}\) & \(3.19{\times}10^{-3}\;(3072)\)
& \(2.94{\times}10^{-2}\;(3072)\) & \(1.22{\times}10^{-2}\;(768)\)\\
17 & 0.16 & \(3.58{\times}10^{-1}\) & \(\mathbf{1.54{\times}10^{-2}}\;(3072)\)
& \(5.71{\times}10^{-1}\;(512)\) & \(6.40{\times}10^{-2}\;(512)\)\\
\bottomrule
\end{tabular}
\end{table}

\vspace{-\baselineskip}
Even after method-specific post hoc selection of \(T\), Greedy spectral has
the smallest mean through \(R=0.04\) in both families.

\begin{table}[H]
\centering
\scriptsize
\setlength{\tabcolsep}{4.2pt}
\caption{Full-state acquisition summary over the four local cells
\(R\le0.04\).  Adaptive ranges include the method-specific selected schedules in Table
\ref{tab:full-state-round-sensitivity}; cost ratios use
\(s:a:r=1:10000:400000\); each adaptive round uses two settings, and ratio
endpoints are rounded upward to integers.}
\label{tab:full-state-acquisition-summary}
\begin{tabular}{@{}cccc@{}}
\toprule
\(d\) & Fixed \(K_+\) & Adaptive rounds \(T\) &
\(C_{\rm adaptive}/C_{\rm fixed}\)\\
\midrule
6  & \(6\)  & \(512\)--\(3072\) & \(1008\)--\(6042\)\\
17 & \(15\) & \(512\)--\(3072\) & \(709\)--\(4251\)\\
\bottomrule
\end{tabular}
\end{table}
\vspace{-\baselineskip}
Across the 24 adaptive comparisons in these eight local cells, every mean infidelity is higher, and every paired interval excludes zero.  Hence the local full-state experiment also gives a joint advantage: lower mean error with zero feedback and adaptive-to-fixed acquisition-cost ratios of
\(709\)--\(6042\).
At \(R=0.08\), S-PAQT has the smaller mean for \(d=6\), with a paired
interval excluding zero; for \(d=17\), fixed spectral has the smaller mean and
the interval contains zero.  At \(R=0.16\), S-PAQT is decisively lower in both
families.
Thus Greedy spectral acquisition leads in both accuracy and acquisition cost in
the calibrated local cells.  In the broadest cells its acquisition advantage
remains, but the adaptive-Bayesian pipeline leads in accuracy.

The supplementary materials separate truth displacement from the assumed
radius for both task and full-state loss, showing how each affects the
crossover: departure from the pilot is the dominant failure mode, while an
overly broad assumed radius accelerates the breakdown of the fixed local
estimator.  Appendix \ref{app:benchmark-contracts} records the exact
benchmark specification and key stability checks.

\noindent\textbf{Code availability.}
Code for reproducing all experiments is available in the
\href{https://github.com/guyhay94/task-directed-quantum-tomography}{https://github.com/guyhay94/task-directed-quantum-tomography}.

\noindent\textbf{Declaration on the use of artificial intelligence.}
OpenAI Codex was used as an assistive tool in drafting and editing portions
of the manuscript and in developing and checking mathematical arguments and
proofs. The authors independently reviewed and verified all mathematical
claims, derivations, proofs, algorithms, citations, code, computational
results, and figures. The authors assume responsibility for all content.

\appendix

\section{Benchmark Specification and Numerical Stability}
\label{app:benchmark-contracts}
\small

\subsection{Benchmark specification}

For the task benchmark, sites are numbered \(i=0,\ldots,5\), and the nominal
Hamiltonian and ordered perturbation library in \eqref{eq:ising-family} are
\[
\resizebox{\linewidth}{!}{$H_0 = -\sum_{i=0}^{4}Z_iZ_{i+1} - 0.97\sum_{i=0}^{5}X_i - 0.08\sum_{i=0}^{5}\sin\{1.37(i+1)\}Z_i, \quad \{V_j\} = 0.35\{X_0,\dots,X_5,Z_0,\dots,Z_5,Z_0Z_1,\dots,Z_4Z_5\}.$}
\]
The Schmidt-spectrum task uses the odd--even subsystem
\(A=\{0,2,4\}\) and complement \(A^c=\{1,3,5\}\).
Unscaled Pauli strings have unit root-mean-square operator norm.  The
structured uniform-ball cloud supplies S-PAQT particles; covariance
matching gives precisions \(20000,35000,42500\) in local dimensions 6, 12, and
15.

S-PAQT initializes its adaptive central iteration in the common pilot and uses
independent Rademacher perturbations with
\(\alpha_k=10/k^{0.602}\) and \(\epsilon_k=0.1/k^{0.101}\).  Its endpoints are
the Bayesian mean state for full-state loss and posterior mean particle
spectrum for task loss.  Particle weights are updated after every count block;
ESS below half the particle count triggers Liu--West resampling with
\(a=0.98\), with out-of-ball proposals rejected.  S-SGQT and S-OSGQT use gains \(\alpha=0.05,\beta=0.2\); OSGQT
also subtracts the simulator-fidelity difference.  Sampling disclosure:
iterates and S-PAQT support are ball-constrained, but paired probes are
unclipped; clipping would break the symmetric difference and can bias its
gradient estimate.  The primary task comparison
uses 8000 S-PAQT particles and 150, 100, 50, or 25 shots per setting.  Radius
diagnostics use 3000 uniform-ball draws for confidence-support volume
and 500 particles.  Broad-radius rankings are exploratory.

\subsection{Numerical stability}

The primary \(R=0.02\) benchmark uses centered finite-difference step
\(10^{-2}\) and relative metric and task cutoffs \(10^{-8}\) and
\(10^{-9}\).  For \(d=17\), the metric cutoff removes the rescaling mode
\((1.09\times10^{-15})\) and one near-null mode
\((6.21\times10^{-9}<1.79\times10^{-8})\), leaving effective dimension 15.
Steps \(0.005\) and \(0.02\) retain dimension 15, rank seven, and the same two
probes; at \(B=153600\), allocations change by $\le14$ shots
(\(<0.01\%\)).

\bibliographystyle{siamplain}
\bibliography{references}

@misc{ferrie_sgqt_2014,
  author = {Ferrie, Christopher},
  title = {Self-Guided Quantum Tomography},
  year = {2014},
  eprint = {1406.4101},
  archivePrefix = {arXiv},
  primaryClass = {quant-ph},
  doi = {10.48550/arXiv.1406.4101}
}

@misc{dekkers_osgqt_2026,
  author = {Dekkers, Kiki and Ruget, Alice and Nothlawala, Fazilah and Henry, Sabrina and Scholes, Stirling and Padgett, Miles and Forbes, Andrew and Nape, Isaac and Leach, Jonathan},
  title = {Orthogonalised Self-Guided Quantum Tomography: Insights from Single-Pixel Imaging},
  year = {2026},
  eprint = {2604.08057},
  archivePrefix = {arXiv},
  primaryClass = {quant-ph}
  }

@misc{granade_paqt_2016,
  author = {Granade, Christopher and Ferrie, Christopher and Flammia, Steven T.},
  title = {Practical Adaptive Quantum Tomography},
  year = {2016},
  eprint = {1605.05039},
  archivePrefix = {arXiv},
  primaryClass = {quant-ph}
  }

@article{stokes_quantum_natural_gradient,
  author = {Stokes, James and Izaac, Josh and Killoran, Nathan and Carleo, Giuseppe},
  title = {Quantum Natural Gradient},
  journal = {Quantum},
  volume = {4},
  pages = {269},
  year = {2020},
  doi = {10.22331/q-2020-05-25-269}
}

@misc{hay_sharon_anchor_projection,
  author = {Hay, Guy and Sharon, Nir},
  title = {Anchor-Based Function Extrapolation with Proven Bounds and Projection Guarantees},
  year = {2026},
  eprint = {2603.09937},
  archivePrefix = {arXiv},
  primaryClass = {math.NA}
  }

@misc{hay_sharon_next_manifolds,
  author = {Hay, Guy and Sharon, Nir},
  title = {Function Extrapolation with Neural Networks and Its Application for Manifolds},
  year = {2024},
  eprint = {2405.10563},
  archivePrefix = {arXiv},
  primaryClass = {cs.LG}
  }

@book{golub_van_loan_matrix_computations,
  author = {Golub, Gene H. and Van Loan, Charles F.},
  title = {Matrix Computations},
  edition = {4th},
  publisher = {Johns Hopkins University Press},
  address = {Baltimore},
  year = {2013}
}

@article{chaloner_verdinelli_bayesian_design,
  author = {Chaloner, Kathryn and Verdinelli, Isabella},
  title = {Bayesian Experimental Design: A Review},
  journal = {Statistical Science},
  volume = {10},
  number = {3},
  pages = {273--304},
  year = {1995},
  doi = {10.1214/ss/1177009939}
}

@article{huan_marzouk_simulation_design,
  author = {Huan, Xun and Marzouk, Youssef M.},
  title = {Simulation-Based Optimal Bayesian Experimental Design for Nonlinear Systems},
  journal = {Journal of Computational Physics},
  volume = {232},
  number = {1},
  pages = {288--317},
  year = {2013},
  doi = {10.1016/j.jcp.2012.08.013}
}

@book{pukelsheim_optimal_design,
  author = {Pukelsheim, Friedrich},
  title = {Optimal Design of Experiments},
  series = {Classics in Applied Mathematics},
  volume = {50},
  publisher = {Society for Industrial and Applied Mathematics},
  address = {Philadelphia},
  year = {2006},
  doi = {10.1137/1.9780898719109}
}

@article{harman_filova_richtarik_exchange,
  author = {Harman, Radoslav and Filov\'{a}, Lenka and Richt\'{a}rik, Peter},
  title = {A Randomized Exchange Algorithm for Computing Optimal Approximate Designs of Experiments},
  journal = {Journal of the American Statistical Association},
  volume = {115},
  number = {529},
  pages = {348--361},
  year = {2020},
  doi = {10.1080/01621459.2018.1546588}
}

@book{nielsen_chuang_qcqi,
  author = {Nielsen, Michael A. and Chuang, Isaac L.},
  title = {Quantum Computation and Quantum Information},
  edition = {10th Anniversary},
  publisher = {Cambridge University Press},
  address = {Cambridge},
  year = {2010}
}

@article{aaronson_shadow_tomography,
  author = {Aaronson, Scott},
  title = {Shadow Tomography of Quantum States},
  journal = {SIAM Journal on Computing},
  volume = {49},
  number = {5},
  pages = {368--394},
  year = {2020},
  doi = {10.1137/18M120275X},
  eprint = {1711.01053},
  archivePrefix = {arXiv},
  primaryClass = {quant-ph}
}

@article{huang_kueng_preskill_classical_shadows,
  author = {Huang, Hsin-Yuan and Kueng, Richard and Preskill, John},
  title = {Predicting Many Properties of a Quantum System from Very Few Measurements},
  journal = {Nature Physics},
  volume = {16},
  pages = {1050--1057},
  year = {2020},
  doi = {10.1038/s41567-020-0932-7},
  eprint = {2002.08953},
  archivePrefix = {arXiv},
  primaryClass = {quant-ph}
}

@article{buhrman_quantum_fingerprinting,
  author = {Buhrman, Harry and Cleve, Richard and Watrous, John and de Wolf, Ronald},
  title = {Quantum Fingerprinting},
  journal = {Physical Review Letters},
  volume = {87},
  number = {16},
  pages = {167902},
  year = {2001},
  doi = {10.1103/PhysRevLett.87.167902}
}

@article{li_haldane_entanglement_spectrum,
  author = {Li, Hui and Haldane, F. D. M.},
  title = {Entanglement Spectrum as a Generalization of Entanglement Entropy: Identification of Topological Order in Non-Abelian Fractional Quantum Hall Effect States},
  journal = {Physical Review Letters},
  volume = {101},
  number = {1},
  pages = {010504},
  year = {2008},
  doi = {10.1103/PhysRevLett.101.010504},
  eprint = {0805.0332},
  archivePrefix = {arXiv},
  primaryClass = {cond-mat.mes-hall}
}

@article{harty_high_fidelity_2014,
  author = {Harty, T. P. and Allcock, D. T. C. and Ballance, C. J. and Guidoni, L. and Janacek, H. A. and Linke, N. M. and Stacey, D. N. and Lucas, D. M.},
  title = {High-Fidelity Preparation, Gates, Memory, and Readout of a Trapped-Ion Quantum Bit},
  journal = {Physical Review Letters},
  volume = {113},
  number = {22},
  pages = {220501},
  year = {2014},
  doi = {10.1103/PhysRevLett.113.220501},
  eprint = {1403.1524},
  archivePrefix = {arXiv},
  primaryClass = {quant-ph}
}

@misc{leu_polarisation_insensitive_2024,
  author = {Leu, A. D. and Smith, M. C. and Gely, M. F. and Lucas, D. M.},
  title = {Polarisation-Insensitive State Preparation for Trapped-Ion Hyperfine Qubits},
  year = {2024},
  eprint = {2406.14448},
  archivePrefix = {arXiv},
  primaryClass = {quant-ph}
  }

@misc{kosut_walmsley_rabitz_oed,
  author = {Kosut, Robert L. and Walmsley, Ian A. and Rabitz, Herschel},
  title = {Optimal Experiment Design for Quantum State and Process Tomography and Hamiltonian Parameter Estimation},
  year = {2004},
  eprint = {quant-ph/0411093},
  archivePrefix = {arXiv},
  primaryClass = {quant-ph},
  doi = {10.48550/arXiv.quant-ph/0411093}
}

@article{nunn_optimal_tomography,
  author = {Nunn, J. and Smith, B. J. and Puentes, G. and Lundeen, J. S. and Walmsley, I. A.},
  title = {Optimal Experiment Design for Quantum State Tomography: Fair, Precise, and Minimal Tomography},
  journal = {Physical Review A},
  volume = {81},
  number = {4},
  pages = {042109},
  year = {2010},
  doi = {10.1103/PhysRevA.81.042109}
}

@article{huszar_houlsby_adaptive_bayesian,
  author = {Husz{\'a}r, Ferenc and Houlsby, Neil M. T.},
  title = {Adaptive Bayesian Quantum Tomography},
  journal = {Physical Review A},
  volume = {85},
  number = {5},
  pages = {052120},
  year = {2012},
  doi = {10.1103/PhysRevA.85.052120}
}

@misc{ito_latency_aware_2023,
  author = {Ito, Kosuke},
  title = {Latency-Aware Adaptive Shot Allocation for Run-Time Efficient Variational Quantum Algorithms},
  year = {2023},
  eprint = {2302.04422},
  archivePrefix = {arXiv},
  primaryClass = {quant-ph}
}

\end{document}